\documentclass[twocolumn,times,tighten]{aastex61}

\usepackage{graphicx}		
\usepackage{latexsym}		
\usepackage{bm}			
\usepackage[english]{babel}
\usepackage{booktabs}	
\usepackage{multirow}	
\usepackage{epstopdf}
\usepackage{hyperref}
\usepackage{cleveref}

\newcommand{\angstrom}{\textup{\AA}}

\newcommand{\U}{\emph{U}}
\newcommand{\Ub}{\emph{U}-band }%
\newcommand{\mab}{$m_{AB}$ }
\newcommand{\sbab}{$\mu^{AB}_{U}$ }
\newcommand{\Uab}{$U_{AB}$ }
\newcommand{\mabu}{$m^{AB}_{U}$ }

\newcommand{\B}{\emph{B}}

\newcommand{\Usp}{$U_{\rm spec}$ }
\newcommand{\sextractor}{\textsc{SExtractor} }
\newcommand{\idl}{\textsc{IDL} }

\newcommand{\swarp}{\textsc{swarp} }

\newlength{\txw}
\newlength{\txh}

\shorttitle{LBT \Ub GOODS-N: Depth vs.\ Resolution}
\shortauthors{Ashcraft et al.}
\submitjournal{Submitted to PASP; Open to comments from the community}

\begin{document}

\title{Ultra-deep Large Binocular Camera \Ub Imaging of the GOODS-North Field: Depth vs.\ Resolution\footnote{Based on data acquired using the Large Binocular Telescope (LBT)}}

\correspondingauthor{Teresa A. Ashcraft}
\email{teresa.ashcraft@asu.edu}

\author{Teresa A. Ashcraft}
\affiliation{School of Earth and Space Exploration, Arizona State University, Tempe, AZ 85287-1404, USA}
\author{Rogier A. Windhorst}
\affiliation{School of Earth and Space Exploration, Arizona State University, Tempe, AZ 85287-1404, USA}
\author{Rolf A. Jansen}
\affiliation{School of Earth and Space Exploration, Arizona State University, Tempe, AZ 85287-1404, USA}
\author{Seth H. Cohen}
\affiliation{School of Earth and Space Exploration, Arizona State University, Tempe, AZ 85287-1404, USA}

 \author{Andrea Grazian}
 \affiliation{INAF - Osservatorio Astronomico di Roma, via Frascati 33, I-00078 Monte Porzio Catone, Italy}
\author{Diego Paris}
\affiliation{INAF - Osservatorio Astronomico di Roma, via Frascati 33, I-00078 Monte Porzio Catone, Italy}
\author{Adriano Fontana} 
\affiliation{INAF - Osservatorio Astronomico di Roma, via Frascati 33, I-00078 Monte Porzio Catone, Italy}
\author{Emanuele Giallongo}
 \affiliation{INAF - Osservatorio Astronomico di Roma, via Frascati 33, I-00078 Monte Porzio Catone, Italy}
\author{Roberto Speziali} 
\affiliation{INAF - Osservatorio Astronomico di Roma, via Frascati 33, I-00078 Monte Porzio Catone, Italy}
\author{Vincenzo Testa}
\affiliation{INAF - Osservatorio Astronomico di Roma, via Frascati 33, I-00078 Monte Porzio Catone, Italy}
 
\author{Konstantina Boutsia}
\affiliation{Carnegie Observatories, Las Campanas Observatory, Colina El Pino, Casilla 601, La Serena, Chile}

\author{Robert W. O'Connell} 
\affiliation{Department of Astronomy, University of Virginia, Charlottesville, VA 22904-4325, USA}

\author{Michael J. Rutkowski} 
\affiliation{Stockholm University, Department of Astronomy \& Oskar Klein Centre for Cosmoparticle Physics, AlbaNova University Centre, SE-106 91 Stockholm, Sweden}

\author{Russell E. Ryan}
\affiliation{Space Telescope Science Institute, Baltimore, MD 21218, USA}

 \author{Claudia Scarlata}
\affiliation{Minnesota Institute for Astrophysics, University of Minnesota, 116 Church Street SE, Minneapolis, MN 55455, USA}
 
\author{Benjamin Weiner}
\affiliation{Steward Observatory, 933 N. Cherry St., University of Arizona, Tucson, AZ 85721, USA}

\begin{abstract}
We present a study of the trade-off between depth and resolution using a large number of \Ub imaging observations in the GOODS-North field (Giavalisco et al. 2004) from the Large Binocular Camera (LBC) on the Large Binocular Telescope (LBT). Having acquired over 30 hours of data (315 images with 5-6 mins exposures), we generated multiple image mosaics, starting with the best atmospheric seeing images (FWHM $\lesssim0\farcs8$), which constitute $\sim$10\% of the total data set. For subsequent mosaics, we added in data with larger seeing values until the final, deepest mosaic included all images with FWHM $\lesssim1\farcs8$ ($\sim$94\% of the total data set). From the mosaics, we made object catalogs to compare the optimal-resolution, yet shallower image to the lower-resolution but deeper image. We show that the number counts for both images are $\sim$90\% complete to \Uab $\lesssim26$. Fainter than \Uab$\sim$\,27, the object counts from the optimal-resolution image start to drop-off dramatically (90\% between \Uab = 27 and 28 mag), while the deepest image with better surface-brightness sensitivity (\sbab$\lesssim$ 32 mag arcsec$^{-2}$) show a more gradual drop (10\% between \Uab $\simeq$ 27 and 28 mag). 

For the brightest galaxies within the GOODS-N field, structure and clumpy features within the galaxies are more prominent in the optimal-resolution image compared to the deeper mosaics. To further investigate how the seeing conditions affect the mosaics, we combined the images by weighting based on the image FWHM. In these weighted stacks, a larger number of small faint galaxies are detected. We conclude that for studies of brighter galaxies and features within them, the optimal-resolution image should be used. However, to fully explore and understand the faintest objects, the deeper imaging with lower resolution are also required. Finally, we find --- for 220 brighter galaxies with \Uab$\lesssim$ 24 mag --- only marginal differences in total flux between the optimal-resolution and lower-resolution light-profiles to \sbab$\lesssim$ 32 mag arcsec$^{-2}$. In only 10\% of the cases are the total-flux differences larger than 0.5 mag. This helps constrain how much flux can be missed from galaxy outskirts, which is important for studies of the Extragalactic Background Light.

\end{abstract}


\section{INTRODUCTION}

In the past 15 years of operation of the \textit{Hubble} and \textit{Chandra} space telescopes, a handful of regions of the sky have been studied to probe the distant universe over relatively wide fields with the aim of understanding the assembly of faint galaxies (e.g., the Cosmic Assembly Near-infrared Deep Extragalactic Legacy Survey, ``CANDELS": Cosmic Evolution Survey, ``COSMOS"; UKIDSS Ultra-Deep Survey, ``UDS"; Extended Groth Strip, ``EGS"; Great Observatories Origins Deep Survey, ``GOODS"; Grogin et al. 2011; Koekemoer et al. 2011). The GOODS field (Giavalisco et al. 2004) contains the deepest data on the sky from many telescopes: \textit{Chandra} (Brandt et al. 2001; Alexander et al. 2003; Xue et al. 2016), \textit{XMM-Newton} (Comastri et al. 2011), \textit{Hubble}, \textit{Spitzer} (Teplitz et al. 2005, 2011; Frayer et al. 2006), \textit{Herschel} (Elbaz et al. 2011), the Very Large Array (VLA; Morrison et al. 2010), the Very Large Telescope (VLT) deep \emph{K}-band survey HUGS (Fontana et al. 2014) and other observatories both in space and from the ground. Together, the GOODS-North and GOODS-South fields subtend $\sim$\,320 arcmin$^2$. The GOODS-N field is centered near RA = 12h 37m, DEC = +62$\deg$ 15' (J2000) and has been observed across the electromagnetic spectrum from radio to X-rays. \textit{HST} has imaged the GOODS-N field from \emph{B} (F435W) to \emph{H} (F160W) at $0\farcs08$ to $0\farcs19$ FWHM resolution and using $0\farcs03-0\farcs06$ pixels in the mosaics (Giavalisco et al. 2004; Grogin et al. 2011; Koekemoer et al. 2011). In the central region of the GOODS-N field, HST UV imaging of F275W and F336W is available with the reduced data having a $0\farcs06$ pixel scale, achieving AB$\,\simeq27.5$ mag $5\sigma$-sensitivity for point sources (\textit{HST} Program: 13872 PI: Oesch; Grogin et al. 2011).

Deep \Ub imaging from the ground can complement the far more expensive HST near-UV imaging. The \Ub ($\lambda_c\,\simeq359$ nm; $\Delta\lambda\simeq54$ nm) is the shortest wide bandpass that can be readily observed from the ground, since the atmosphere is opaque below $\sim\,$320 nm. Most ground-based optical telescopes have some \Ub capabilities, but many CCDs and camera optics do not optimally perform at these near-UV wavelengths. The importance of \Ub observations has led some of the largest telescopes in the world (e.g., the Large Binocular Telescope (LBT); the Very Large Telescopes (VLT); the Subaru Telescope) to include instruments that can observe efficiently at these near-UV wavelengths. 

Matching HST resolution is not possible from the ground, but using images with the best seeing conditions can minimize the impact of image blurring. Unlike in space, on the ground --- even at the best locations --- the seeing conditions vary significantly during each night and with wavelength. Taylor et al.\,(2004) give an overview of seeing conditions on Mt. Graham where the LBT is located. Therefore, when observing for multiple nights, the seeing (FWHM), as measured in the data, will also significantly vary.

There are several \Ub surveys of multi-wavelength HST and other deep fields. A previous \Ub survey of GOODS-N includes the 4m KPNO survey with a $5\sigma$ limit of \mabu = 27.1 mag  and with $1\farcs26$ FWHM seeing (Capak et al. 2004). Grazian et al. (2009) used the Large Binocular Camera (LBC) on the LBT, and derived the galaxy number counts with a 30\% completeness level in the \Ub (\U-Bessel + \Usp filters) to \mab = 27.86 mag. They observed 4 different fields under varying seeing conditions ($1\farcs0-1\farcs4$ FWHM) and depths (\mabu= 25.86 -- 27.86 mag). The VLT VIMOS instrument has surveyed the GOODS-S field in the \Ub to depths of \mabu = 29.8 mag ($1\sigma$; or $\sim$\,28.05 mag at $5\sigma$) with a resolution of $0\farcs8$ FWHM (Nonino et al. 2009). 

During its remaining lifetime of HST may observe the remainder ''CANDELS" fields, including GOODS-N in the NUV (225-275 nm). Currently, there is no space-based replacement for observing at these NUV wavelengths once HST becomes inoperable. The LBT is able to get \Ub imaging for 4 of the 5 ``CANDELS" fields at flux limits comparable to what HST can do in the F336W filter. In this paper, we therefore present ultra-deep \Ub imaging of GOODS-N and make mosaics based on optimal-resolution, and optimal-depth to show the best \Ub imaging which currently can be done from the ground. 

The Extragalactic Background Light (EBL; Dwek \& Krennich 2013, and references therein) is the flux received today mainly from star-formation processes in  the extragalactic sky from the far-UV to the far-IR. There are two main types of measurements, ``direct measurements" and ``integrated galaxy counts" which currently find a factor of 3-5 disagreement at the UV -- optical wavelengths (Driver et al. 2016). The reasons for the discrepancies could be due to integrated galaxy counts underestimating flux in the outer-parts of galaxies, \emph{or} perhaps that some of the direct measurements include fore-ground contaminants e.g. zodiacal light (Driver et al. 2016). Our ultra-deep \Ub imaging allows us to explore the surface brightness of galaxies to very faint limits in the context of Extragalactic Background Light (EBL).

This paper is organized as follows. In \S$\,2$, we describe the data acquired from the LBT, and how we made our mosaics and object catalogs. In \S$\,3$, we present our comparison of the optimal-resolution and optimal-depth mosaics. In \S$\,$4, we summarize our results. All magnitudes presented in this paper are in AB (Oke \& Gunn 1983).

\begin{table}
\begin{center}
\caption{\noindent\small LBT/LBCB \emph{U}$\!_{\text{spec}}$ Exposures \label{table:images}} 
\begin{tabular}{cccc}
\tableline\tableline
Group & Number of  &Exposure Time  &Total Exposure\\      
 & Images &Per Image (s) &Time (Hours)\\
\tableline
Italian   & 272 & 360  & 27.22\\
US  & 75  & 300 & 6.25\\

\tableline
\end{tabular}
\end{center}
\end{table}

\section{OBSERVATIONS}
\label{sec:obs}
\subsection{Large Binocular Camera Capabilities}

The Large Binocular Cameras (LBCs; Giallongo et al. 2008) consists of two wide-field prime focus instruments on the LBT, each with a $\sim23.6'\times25.3'$ field of view (FoV), which can be operated simultaneously. Each camera consists of four $4\,\rm{K}\,\times\,2\,\rm{K}$, E2V 42-90 CCDs with a pixel-size of $\sim0\farcs2254$\,pix$^{-1}$, a gain of $\sim$\,2.02\,$e^{-}$/ADU and read-noise of $\sim\,$5.0 ADU. Since the layout of the CCDs within the camera is not a square, the total effective FoV is about 470 arcmin$^2$. Its binocular image mode allows the LBCs to observe the \textit{same} portion of the sky simultaneously in \textit{both} the red and the blue/near-UV with the two separate cameras. The LBC instruments --- one for each 8.4m LBT mirror --- are each optimized to observe in either the blue UV--$R$ bands ($350-650$ nm) or in the red $V$--$Y$ bands ($500-1000$ nm), respectively. The SDT\_$U_{\rm spec}$ filter has a central wavelength at $\lambda_c = 3590$ $\angstrom$ and a band width of 540 $\angstrom$ (FWHM). The peak CCD quantum efficiency is $\sim50\%$ in the SDT\_$U_{\rm spec}$ filter (Giallongo et al. 2008).

\subsection{\Ub Observations of the GOODS-N Field}

LBC observations of the GOODS-N field were carried out in dark time from 2012 December to 2014 January. All our LBT observations were made using binocular image mode. Since by far the most exposures were taken in the SDT\_\Usp filter, the current work presents only the LBC-Blue (LBCB) channel images. In the LBC-Red (LBCR) channel, the SDSS $riz$ filters were use. These will be presented in a future paper that will also contain data from upcoming observing runs, since there were not nearly as many exposure in these filters as in the \U-band. Over 27 hours were contributed from Italian partner time, while the remainder came from a collaboration of US LBT partner institutions (Table \ref{table:images}). Combined, we acquired a total of 335 SDT\_U$_{\rm spec}$ science exposures with a total open-shutter time of $\approx32.5$ hours (117,220 seconds). Due to the LBC's large FOV, only one pointing is needed to cover the entire HST GOODS-N field. We implemented both major and minor dither patterns to fill in the gaps between the LBC CCDs, and to remove cosmic rays and detector defects. The observations were a collaborative effort between the US and Italian LBT partners, and as a result, the pointings do not always perfectly match up, nor are the dither patterns always identical. The total usable survey area is $\sim$0.16 deg$^2$. Each individual image has an exposure time of either 300s or 360s, with the majority being the latter. Bias frames and twilight sky-flats were taken on most nights for calibration. All individual science images were reduced using the LBC pipeline as described in Giallongo et al. (2008), which includes bias-subtraction, flat-fielding, and astrometric corrections.

\noindent\begin{figure}[t!]
\includegraphics[width=0.48\txw]{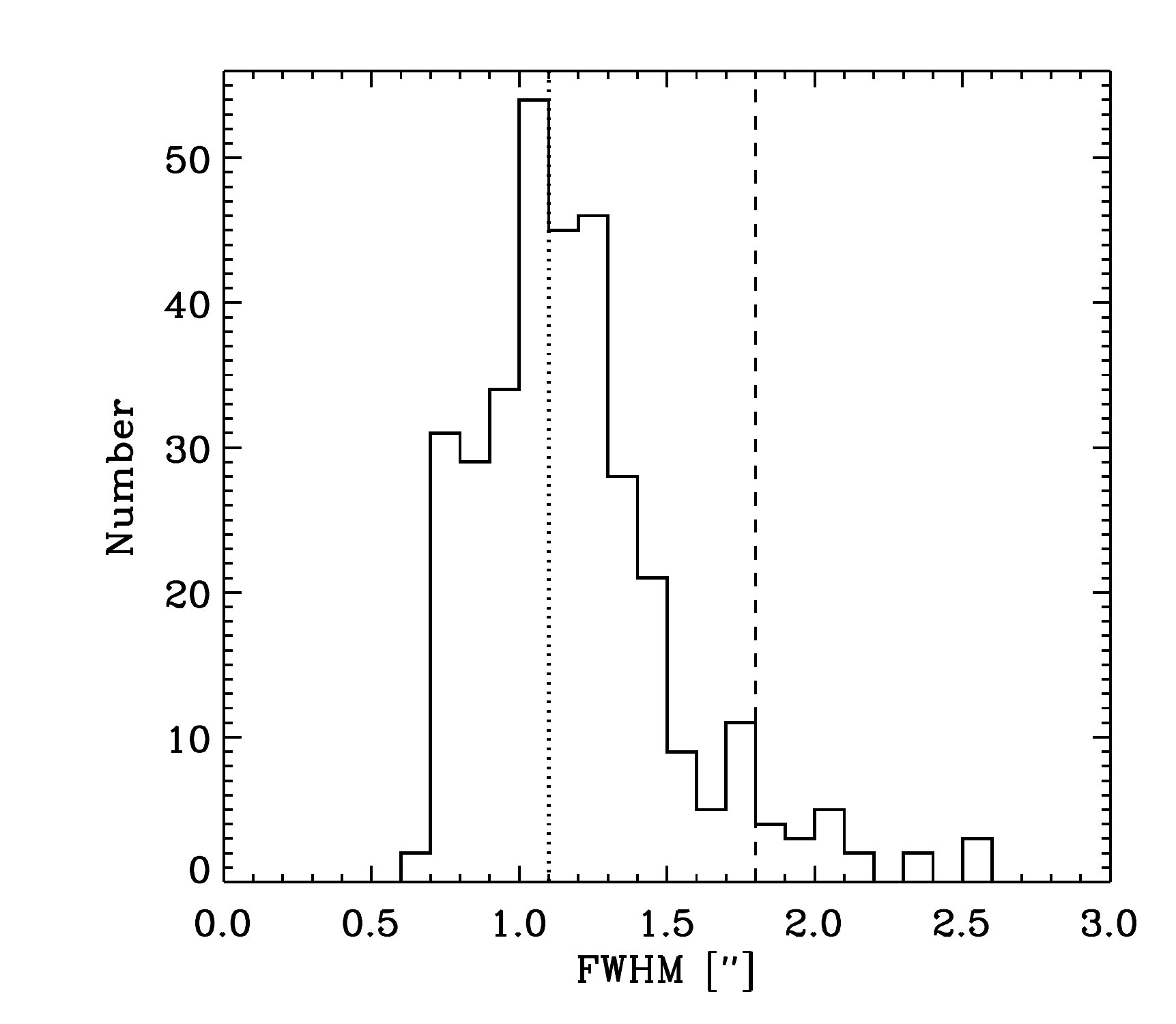}
\caption{\noindent\small
A histogram of the FWHM measured from stars for the 335 individual \Ub exposures taken in the GOODS-N field with the LBT. Each exposure results in four separate images, one for each CCD in the detector array. We show the average FWHM across all four chips, which have an average pixel-size of  $\sim0\farcs225$. The vertical dashed line represents the cut-off of FWHM$\,=1\farcs8$ used for our final image stacking, which excluded the 20 worst exposures. The dotted line represents the median seeing, $\sim1\farcs1$ FHWM, for all exposures with FWHM $\leq$ $1\farcs8$. \label{figure:hist}}
\end{figure}
\vspace*{-\baselineskip}

\subsection{Creating \Ub Mosaics}
\label{sec:mosaics}
For all 32.5 hours, the Gaussian FWHM was measured for $\sim$100 unsaturated stars in all individual exposures, as shown in the histogram in Fig.\,\ref{figure:hist}, which has a median FWHM of $\sim1\farcs1$. Images with poor seeing (FWHM $> 1\farcs8$, or $\sim6\%$ of the data, or 20 images), were excluded in the final stacking. Mosaics were made from subsets of the 315 remaining images. We sorted these images in order of increasing seeing FWHM, and stacked all images with seeing $\lesssim0\farcs8$ FWHM (33 exposures, or $\sim10\%$ of the data). Additional stacks were made by increasing included images by $\Delta$FWHM = $0\farcs1$ increments (e.g. FWHM $\lesssim0\farcs9$). The final mosaic included all 315 images with FWHM $\lesssim1\farcs8$.

The images were combined using the \swarp package (Bertin et al. 2002; Bertin 2010). This program uses astrometric solutions to re-sample and co-add all the FITS images. Within \swarp$\!\!$, we had to set several key parameters to optimize the resampling and stacking of the individual images (Table \ref{table:swarp}). \swarp first subtracts the sky-background from each input image. For background determination, we used a ``back\_size" parameter of 256 pixels for the mesh size, and a ``back\_filtersize" of 3. This resamples the input images using the ``LANCZOS3" as the interpolation function. When resampling, the ``LANCZOS3" function preserves the signal with only minor artifacts from image discontinuities. For co-adding the re-sampled images,  ``combine\_type" was set to ``median", which selects for each output-pixel the median of the non-zero weighted and scaled pixel-values. Each mosaic produced by \swarp is the same size (6351 $\times$ 6751 pixels based on the shallowest mosaic), with the same coordinate J2000 RA = 12:36:54.5, Dec = +62:15:41.1 used for the image center and a ``pixelscale\_type" of median. Weight-maps, which are used in making object catalogs, were created by \swarp as well. For making object catalogs and for other analyses, the outer regions of the mosaics with exposure times $\leq\,$3600s were excluded. This exclusion region was determined by the shallowest mosaic, and applied to all mosaics. The exposure limit was set to ensure that only regions with at least 10 separate exposures will be included.

\noindent\begin{deluxetable}{lc}
\tabletypesize{\footnotesize} 
\tablewidth{0pt}
\tablecolumns{2}
\tablecaption{\noindent\small SWARP Configuration File \label{table:swarp}}
\tablehead{
\colhead{Keyword} & \colhead{Value}}
\startdata
COMBINE\_TYPE & Median \\
WEIGHT\_TYPE  & MAP\_WEIGHT \\
PIXELSCALE\_TYPE & Median \\
CENTER (J2000) & 12:36:54.5, +62:15:41.1 \\
IMAGE\_SIZE (pix) & 6351, 6751  \\
RESAMPLING\_TYPE & LANCZOS3 \\
\enddata
\vspace{-0.8cm} 
\end{deluxetable}

\noindent\begin{deluxetable}{lcc}[b!]
\tabletypesize{\footnotesize} 
\tablewidth{0pt}
\tablecolumns{3}
\tablecaption{\noindent\small \textsc{SExtractor} Configuration File \label{table:sext}}
\tablehead{
\colhead{Keyword} & \colhead{Optimized Resolution} & \colhead{Optimized Depth} }
\startdata
DETECT\_MINAREA & 5 & 5\\
DETECT\_THRESH & 1.0 & 1.0\\
ANALYSIS\_THRESH & 1.0 & 1.0\\
DEBLEND\_NTHRESH & 64 & 64\\
DEBLEND\_MINCONT & 0.06 & 0.04\\
WEIGHT\_TYPE  & MAP\_RMS & MAP\_RMS \\
\enddata
\end{deluxetable}


\noindent\begin{figure*}[t!]
\includegraphics[width=0.96\txw]{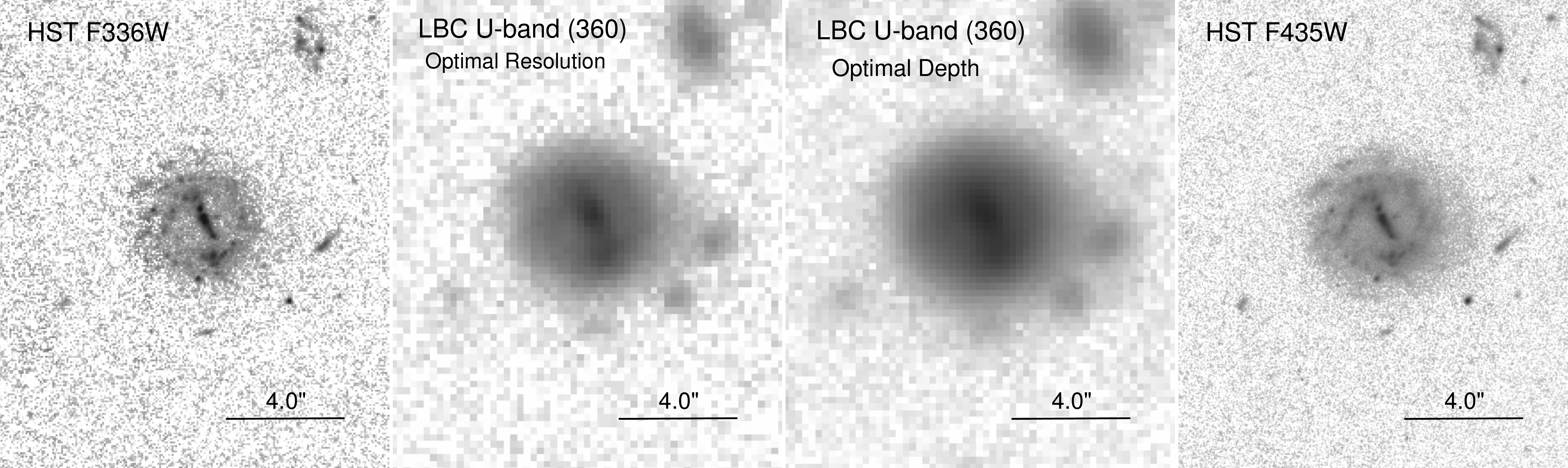}
\includegraphics[width=0.96\txw]{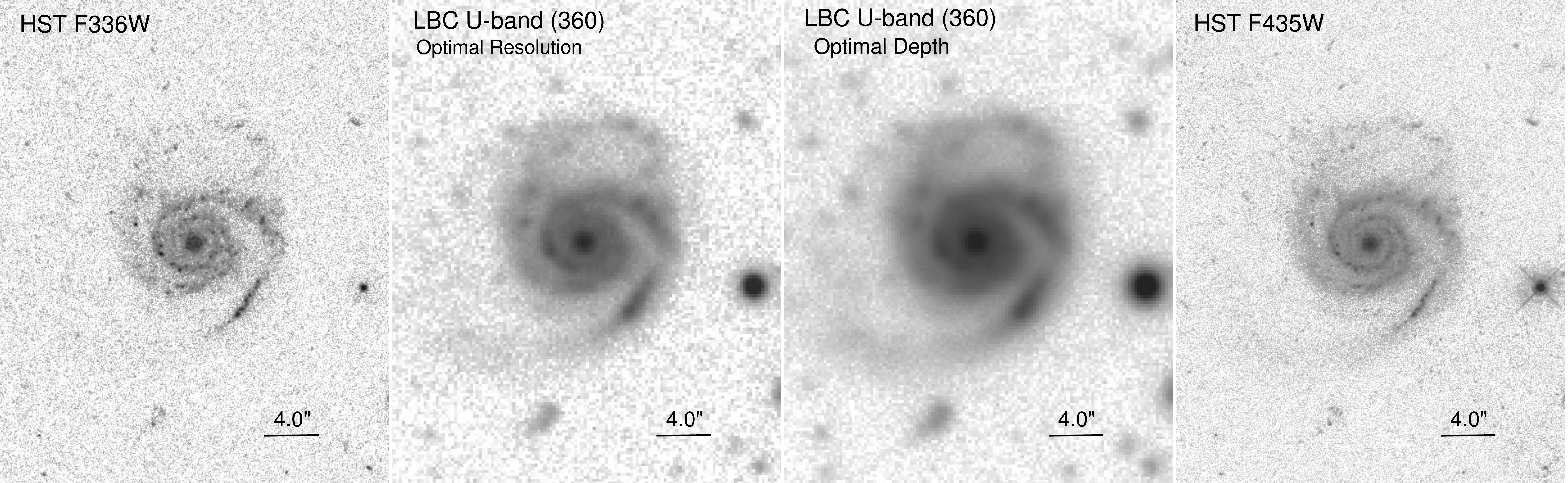}
\caption{\noindent\small
Displayed are two bright galaxies in the GOODS-N field. The top galaxy has \Uab = 20.8 mag, and the bottom galaxy has \Uab = 18.0 mag. Left is the HST WFC3 \Ub (F336W; \textit{HST} Program: 13872 PI: Oesch), middle left is the optimal-resolution \Ub image (3 hrs), middle right is the deepest-lower-resolution \Ub image (30 hrs), and right is the HST ACS $B$-Band (F435W; Giavalisco et al. 2004). To compare our LBT ground-based data to the best existing space-based data (HST) at similar wavelengths, we are showing the only two large, bright galaxies which appear in both the HST \Ub and \B-band. For the top galaxy, the main bar feature seen in the HST images, is also discernible in the LBT mosaics, especially in the optimal-resolution mosaic. Since the bottom galaxy is significantly brighter, there are more galaxy sub-structures visible in the LBT mosaics. Of the detectable features in the LBT mosaics, the smallest galaxy features are easier to identify in the optimal-resolution mosaic. The deeper LBT mosaics better shows extended low-surface brightness flux in the outer parts of the galaxy, which does not stand out or is significantly detected in the HST images, especially in the HST \U-band. \label{figure:brightgals336}}
\end{figure*}

\vspace*{-\baselineskip}
\vspace*{-\baselineskip}
\vspace*{-\baselineskip}

\subsection{LBC \Ub Catalogs}
Object catalogs were made using \sextractor (Bertin et al. 1996). Finding the best combination of \sextractor parameters to both identify faint objects, and to not split brighter extended objects, is a complicated task. For the large-scale sky-background determination, a large mesh of $256\times256$ pixels and a median filter of $3\times3$ pixels were chosen to deal with bright saturated stars and bright, extended galaxies. For local sky-background subtraction, an annulus of 48 pixels was adopted for each object. For object-detection, \sextractor smooths the image using a Gaussian filter with a convolving kernel with a FWHM of 3.0 pixels, and a convolution image size of $5\times5$ pixels. Other parameters that we adapted to optimize were the sigma-limit above the sky-background for initial object detection (1.0$\sigma$), the minimum number of connected pixels (5 pixels), and the deblending parameters ``deblend\_nthresh" and ``deblend\_mincont" (see Table \ref{table:sext}). This allowed us to not break up large objects into multiple detections, yet still distinguish between them in the object catalogs. We refer to \S$\,$\ref{sec:analysis} for the best choice of these parameters for the LBT data.

We generated a mask-image to discard several bright stars and surrounding corrupted areas. The same mask was used for all mosaics, based on the deepest image, which is determined by the larger FWHM of all the unsaturated stars. In the final object catalog, we excluded all objects with the \sextractor parameter flag larger than 3, which are likely defects caused by detection or measurement issues when running \textsc{SExtractor}. Objects with a flag value larger than 3 can be due to a number of  complications, including saturated pixel(s), or may be corrupted by the image boundaries. Table \ref{table:stack} lists the number of images stacked, the maximum seeing FWHM-value of the images included in each stack, and the measured FWHM-value for each final image, as described below.

Photometric zero-points were determined by matching our \sextractor catalogs to the KPNO HDF-N \Ub catalog (Capak et al. 2004). Almost 200 stars with AB-magnitudes between \Uab$\simeq17$ and \Uab$\simeq22$ mag were verified in the LBC image, both visually and by measuring their FWHM. The FWHM-value for each mosaic was measured by averaging the FWHM of these stars. The brightest stars from the KPNO survey were excluded due to saturation in the LBT mosaics. Other stars were missing, because the KPNO survey used the R-band for object detection. To ensure the brightest stars still included were not saturated in individual exposures, the peak flux of stars with AB $\lesssim$ 18 mag were checked, especially for the exposures with the best-seeing, as saturation would most likely be first occur here. All stars checked were well below the saturation level of $\sim$\,65,000 counts. Over 100 non-saturated stars --- found in both survey catalogs --- were used to measure the zero-point for each mosaic. There was a slight shift in the zero-point between the shallowest and deepest image, amounting to $\sim 0.2$ mag, which could indicate transparency differences between individual exposures and between various nights. We refer to Taylor et al. (2004) for a more complete discussion of the seeing, transparency and sky-brightness trends at the Mt. Graham Observatory. To compensate for this zero-point offset, the appropriate zero-point was used when measuring the AB magnitude of objects in each mosaic, i.e., 26.63 mag for the optimal-resolution image, and 26.42 mag for the optimal-depth image. 

 \noindent\begin{figure*}[t]

\includegraphics[width=0.96\txw]{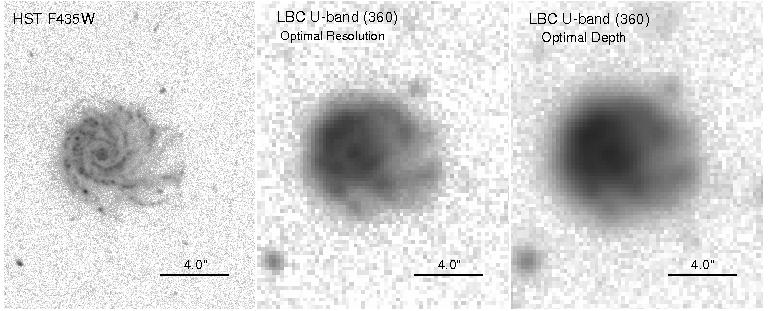}
 \caption{A bright face-on spiral galaxy with \Uab = 19.9 mag. Left is the HST ACS $B$-Band (F435W; Giavalisco et al. 2004), middle is the optimal-resolution \Ub image (3 hrs), and right is the deepest-lower-resolution \Ub image (30 hrs). The HST F336W is not currently available, so for this low redshift galaxy, the closest HST filter in wavelength, F435W, is shown.  While the smaller scale details of the galaxy are only visible in the HST \B-band, some larger scale features are still discernible in the optimal-resolution LBT mosaic. These same large scale features are at best only slightly distinguishable in the optimal-depth mosaic.
\label{figure:brightgals2}}
\end{figure*}

\begin{table}
\begin{center}
\caption{Overview of LBT \emph{U}-band Stacked Mosaics\label{table:stack}} 
\begin{tabular}{cccc}
\tableline\tableline
 Number of  	& Exposure Time	  &    FWHM  & Depth $5\sigma$\\
Stacked Images          &   (Hours)       &  (arcsec) & $U_{AB}$ (Mag)\\              
\tableline
  \phn33 & \phn3.2   & 0.77  & 27.0 \\
 \phn62 & \phn6.0  & 0.81 & 27.4\\
 \phn96 & \phn9.1   & 0.87  & 27.6 \\
 150    & 14.2      & 0.96  & 27.8 \\
 195    & 18.8      & 1.00  & 28.0 \\
 241    & 23.2      & 1.03 & 28.1 \\
 269    & 26.0      & 1.06 & 28.2 \\
 290    & 28.1      & 1.08 & 28.2 \\
 315    & 30.4      & 1.10 & 28.3 \\ 
\tableline
\end{tabular}

\end{center}
\end{table}
\vspace*{-\baselineskip}

\section{ANALYSIS}
\label{sec:analysis}

\begin{figure*}[t]
\includegraphics[width=0.96\txw]{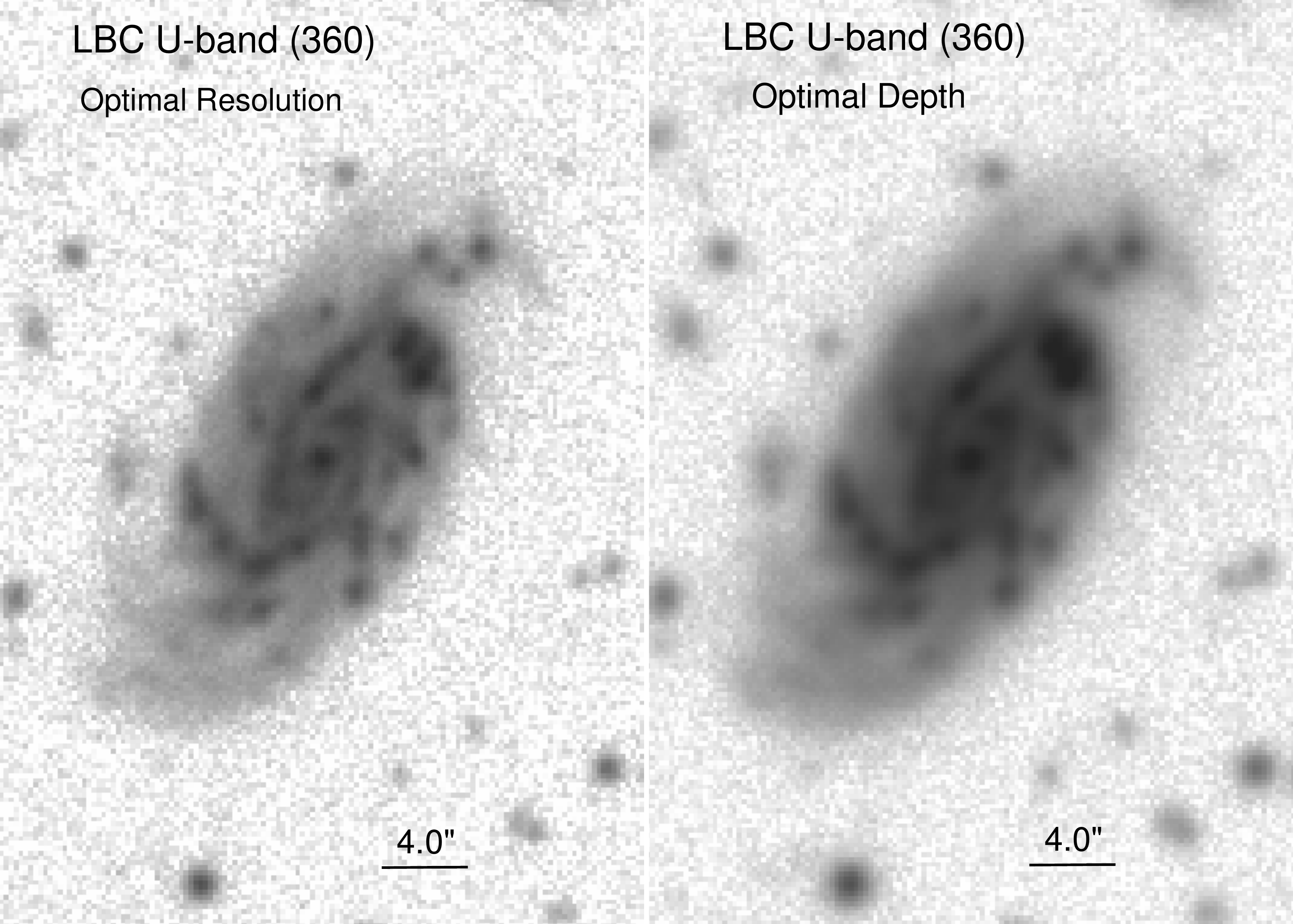}
\includegraphics[width=0.96\txw]{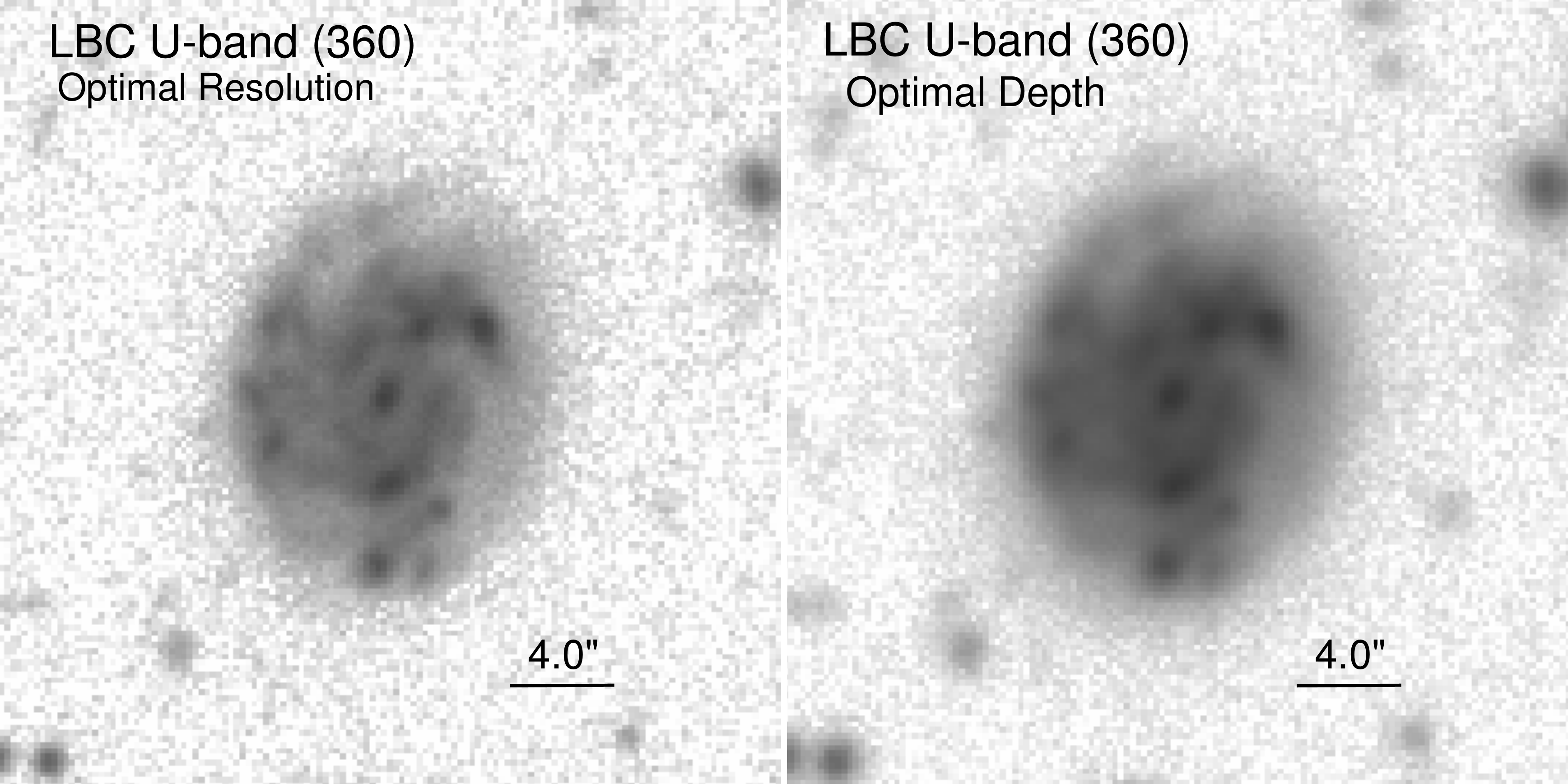}
\caption{ \noindent\small
Two of the brightest galaxies in the LBC FOV, but outside of the HST FOV. The top galaxy has \Uab = 18.3 mag, and the bottom galaxy has \Uab = 19.2 mag. Left is the optimal-resolution \Ub image (3 hrs), and right is the deepest-lower-resolution \Ub image (30 hrs). Both of these face-on spirals are resolved enough to see several sub-structure features including clumps and spiral arms. Most of the features can be seen in both the optimal-resolution and optimal-depth mosaics. However, the features are sharper and easier to distinguish in the optimal-resolution mosaic, while some features blur together in the optimal-depth mosaic.
\label{figure:brightgals3}}
\end{figure*}

\begin{figure*}
\centerline{
\includegraphics[width=0.96\txw]{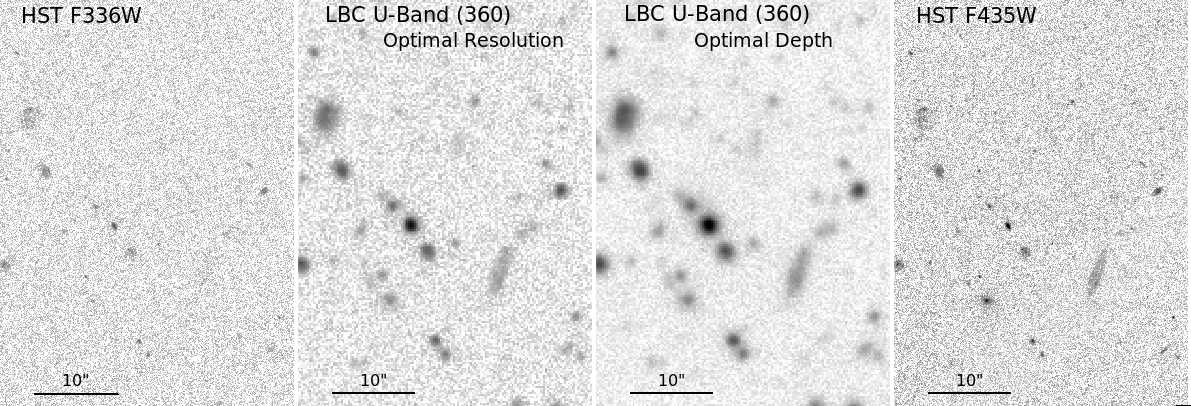}
}
\caption{ \noindent\small
Left is the HST WFC3 $U$-Band (F336W; \textit{HST} Program: 13872 PI: Oesch), middle-left is the optimal-resolution \Ub image (3 hrs), middle-right is the deepest-lower-resolution \Ub image (30 hrs), and right is the HST $B$-Band (F435W; Giavalisco et al. 2004). In this region of the GOODS-N field shown, the faintest objects detected in the LBT \Ub mosaics (\Uab$\sim$27.8 for the optimal-depth image) are not always discernible in the HST images, especially the F336W. The smallest objects in the HST B-band (F435W; FWHM$\lesssim0\farcs7$) are not detected as objects in the LBC mosaic catalogs. \label{figure:faintgals}}
\end{figure*}

\begin{figure}
\centerline{
\includegraphics[width=0.48\txw]{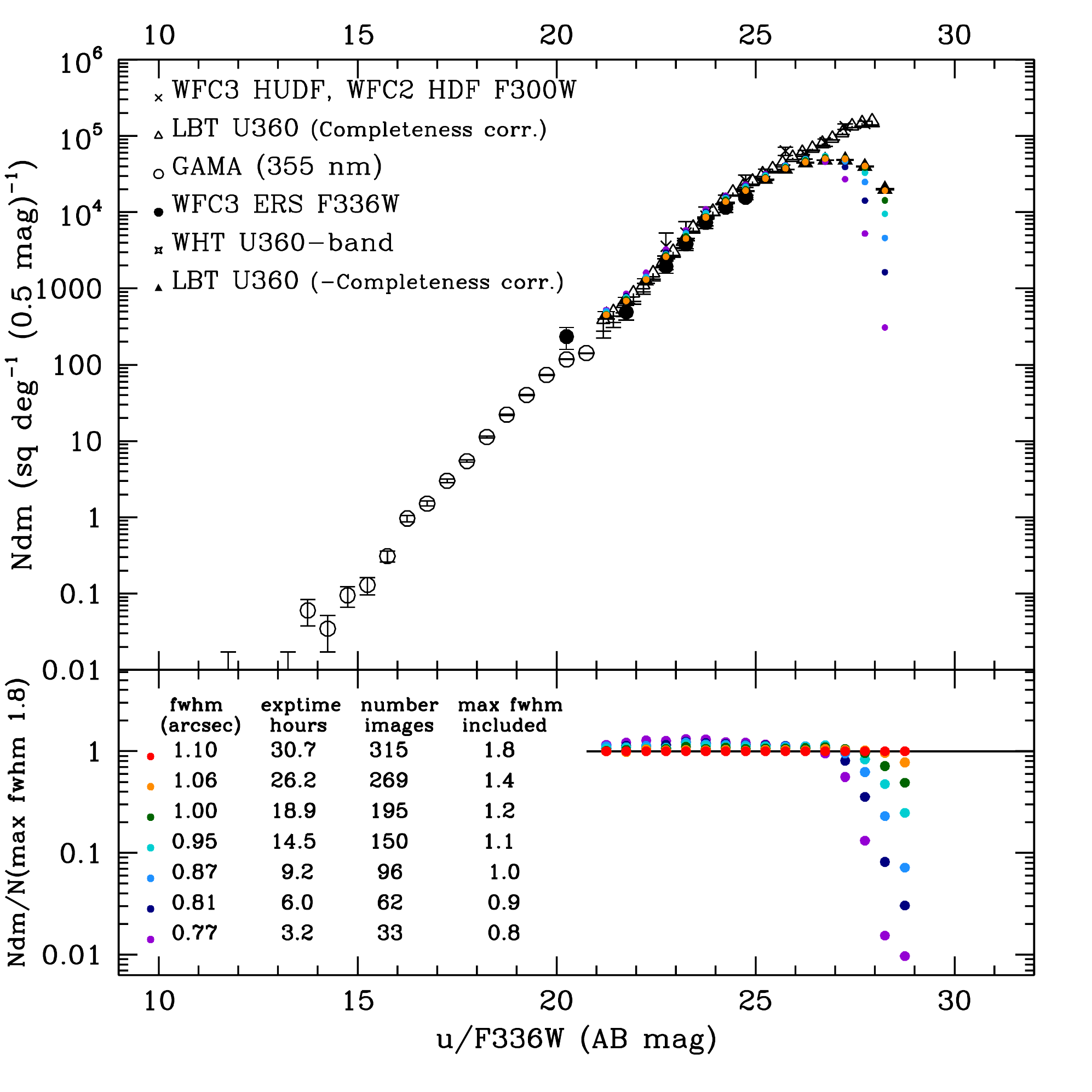}
}
\caption{\noindent\small
Top: Differential number counts for different mosaics compared to other \Ub surveys ( Metcalfe et al. (2001); Driver et al. 2009; Grazian et al. 2009; Windhorst et al. 2011). Bottom: Differential number counts divided by the deepest counts for each mosaic. The orange/red circles are the deepest images, and the purple/blue are the optimal-resolution images. The counts for all mosaics turn over from a power law extrapolation at \Uab$\gtrsim 25.5$ mag. The shallowest optimal-resolution images drop-off dramatically at \Uab$\sim27$ mag, while the fall off is more gradual for deepest images with the lower-resolution. \label{figure:numcnts}}
\end{figure}

When including the lower-resolution images (FWHM $\gtrsim1\farcs0$), the resulting quality of the image degrades, which results in the loss of some clumpy features, especially for larger and brighter galaxies (Fig. \ref{figure:brightgals336} -- Fig. \ref{figure:brightgals3}). This is most apparent when comparing the \Ub $0\farcs8$ and $1\farcs1$ FWHM images to the \textit{HST}-ACS $B_{435}$ (Giavalisco et al. 2004) and \textit{HST}-WFC3 $U_{336}$ (\textit{HST} Program: 13872 PI: Oesch) images of the same bright galaxy (\Uab$\simeq 19-21$ mag) in Fig. \ref{figure:brightgals336}. For Fig. \ref{figure:brightgals2}, the \textit{HST}-ACS $B_{435}$ (Giavalisco et al. 2004) image is shown for comparison, since very few F336W reduced images are available in GOODS-N. The galaxies in Fig.\,\ref{figure:brightgals3} are outside the HST footprint, and so, have no HST imaging to compare to.
The lower-resolution images also make it more difficult to deblend nearby objects (Fig. \ref{figure:faintgals}). 
One way we dealt with this deblending issue was through optimizing the \sextractor parameters, as tabulated in Table 3. For the lower-resolution mosaics, we changed the ``deblend\_mincont" to 0.04, while it was set to 0.06 for the optimal-resolution mosaics. This did not explain the entire difference, and still left a slightly larger number of objects per AB-magnitude bin at brighter fluxes (\Uab\,$\lesssim$\,26 mag) in the optimal-resolution images compared to the lower-resolution number counts (Fig. \ref{figure:numcnts}).

\begin{figure}
\centerline{
\includegraphics[width=0.48\txw]{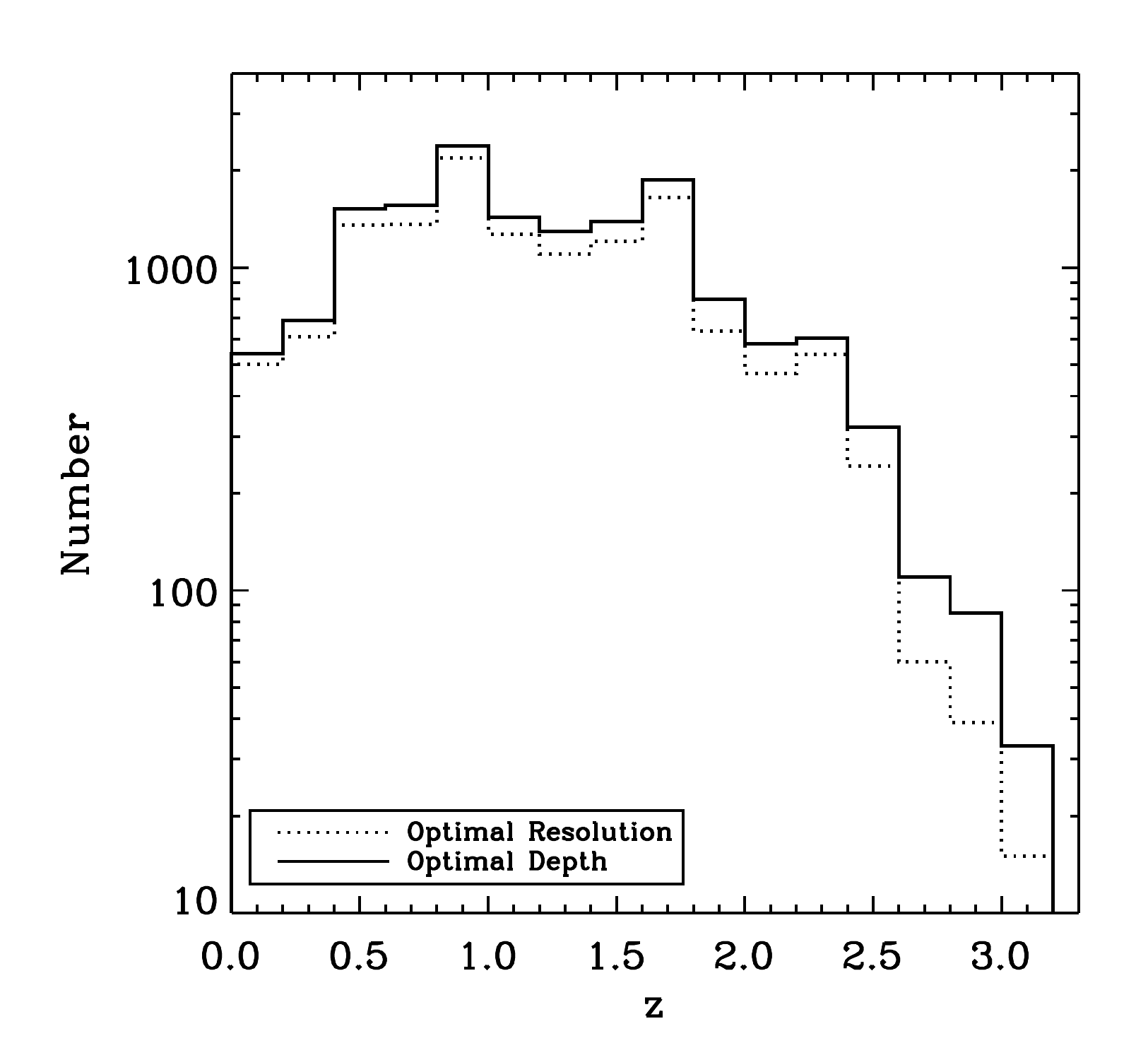}
}
\caption{\noindent\small
Histogram of object redshifts for the deepest (solid) and shallowest (dots) GOODS-N image. Redshifts are a combination of the photometric from 3D HST (Skeleton et al. 2014 and references there in), but using spectroscopic redshifts where available. This shows that more objects are detected the deepest LBT mosaic compare to the shallowest LBT mosaic. The median redshift is z$_{\textrm{med}} \simeq 1.4$, which justifies the use of F435W filter in Fig. 3 when the F336W filter is not available (see Taylor-Mager et al. 2007 for a full discussion). \label{figure:histz}}
\end{figure}

\begin{figure}

\centerline{
\includegraphics[width=0.48\txw]{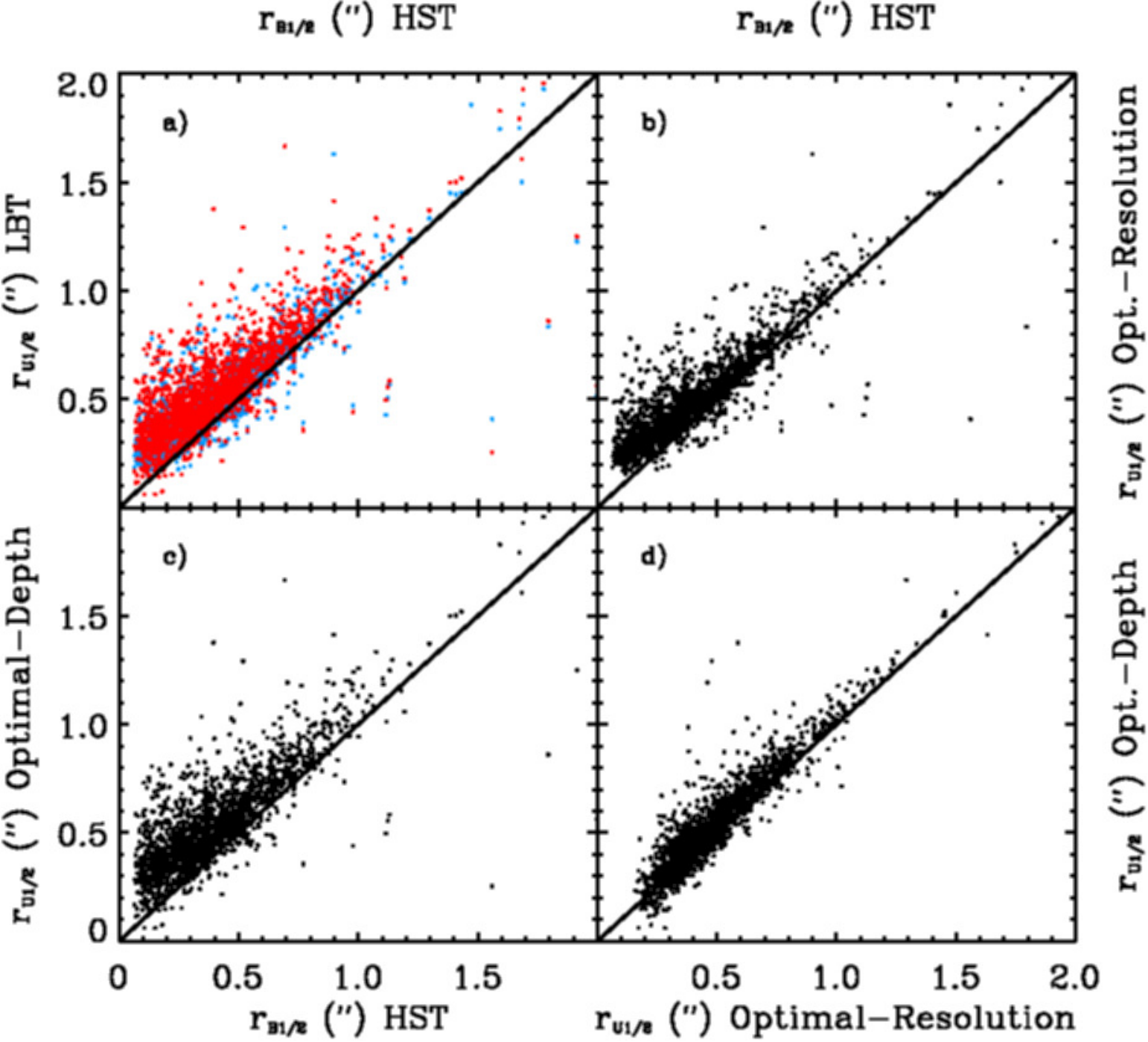}
}
\caption{ \noindent\small
Comparison plots of half-light radius measured by \sextractor for two LBT mosaics, optimal-resolution and optimal-depth, and the HST \B--band. We only included galaxies with \Uab $\leq 25$ mag. All half-light radii plotted have been corrected for FWHM using $r_{corr}=\sqrt{r_{\sextractor}^{2}-(\textrm{FWHM}/2)^{2}}$.  Half-light radius for the optimal-resolution (blue) and lower-resolution (red) images compared to the half-light radius for the HST \B--band (a; Giavalisco et al. 2004). The optimal-resolution (b) or the optimal-depth (c) half-light radius are compared to the HST \B--band. The bottom right panel (d) compares the optimal-depth to the optimal-resolution half-light radius measurements.  \label{figure:rhst}}
\end{figure}

We compared our \sextractor \Ub half-light radii to the equivalent in the $B$-band HST catalogs of the GOODS-N field (Giavalisco et al. 2004). Since there is only currently limited HST \Ub imaging of the GOODS-N field (\textit{HST} Program: 13872 PI: Oesch), we rely on the HST \emph{B}--band images for direct comparison of individual objects. Although these are somewhat different filters, both sample rest-frame wavelengths blueward of the 4000$\angstrom$ break for most of the galaxies at the median redshifts of the sample (see Fig. \ref{figure:histz}), where size differences are less rest-frame wavelength dependent (e.g., Taylor-Mager et al. 2007). We compared objects with $20\leq$\,\mab$\leq 25$ mag as selected in the HST \emph{B}--band. The top left panel of Fig. \ref{figure:rhst} shows that the radii measured in the optimal-resolution image (black dots) agree better with the HST size-measurements with less scatter than the sizes measured in the lower-resolution image (red dots). In order to recover intrinsic object sizes, we subtracted the PSF FWHM-value in quadrature from the best-seeing and the deepest measurements ($0\farcs77$ and, $1\farcs1$ FWHM, resp.). In Fig. \ref{figure:rhst} we show a comparison of the corrected versus uncorrected half-light radius. The PSF-size was subtracted in quadrature for the \emph{B}--band HST images as well, but since the HST/ACS PSF is so small ($0\farcs08$ FWHM; see, Fig. 10a of Windhorst et al. 2011), that this correction had almost no effect, except for the very smallest and faintest objects.

\begin{figure}
\centerline{
\includegraphics[width=0.48\txw]{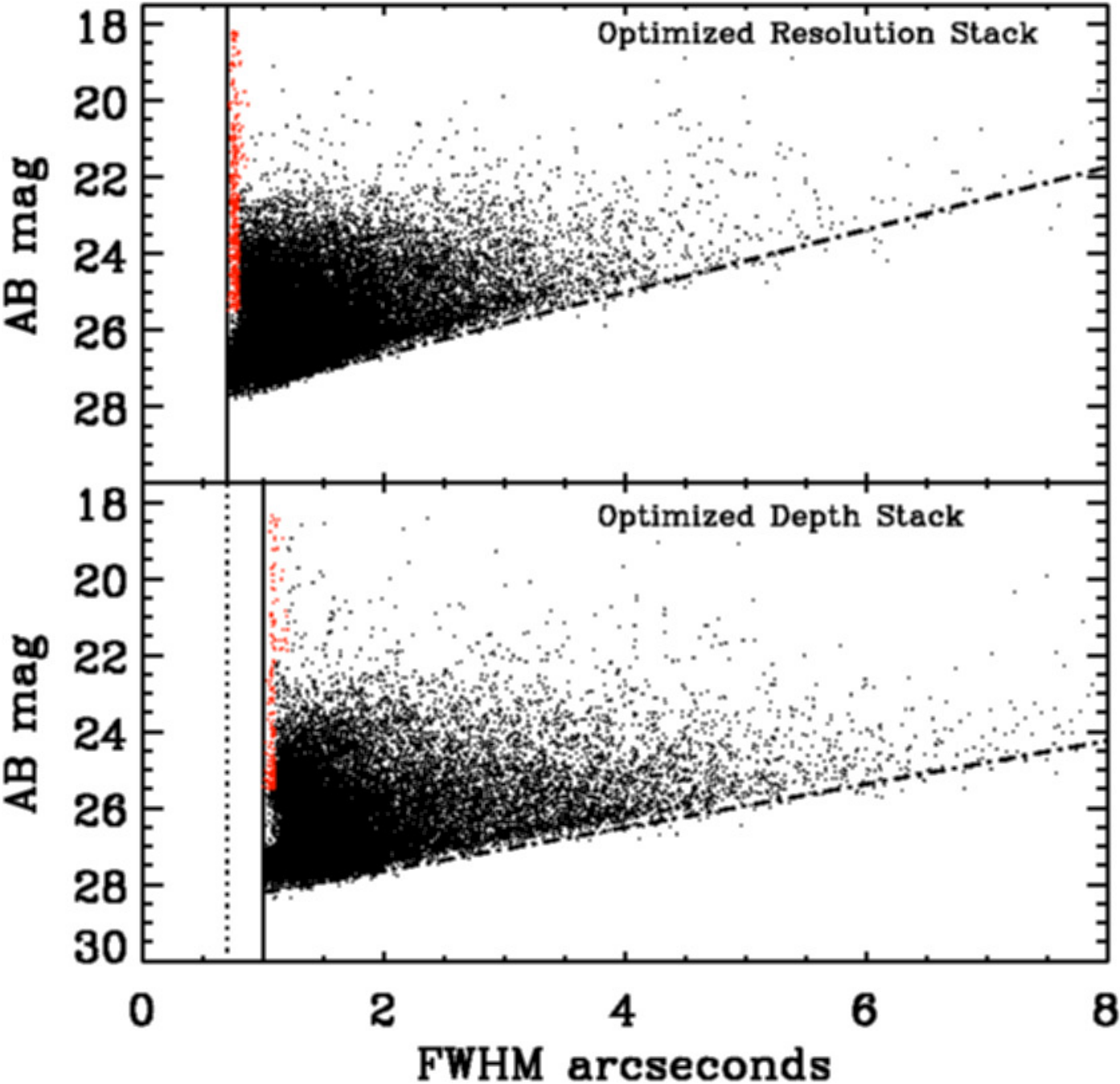}
}
\caption{ \noindent\small
Total \sextractor object magnitude vs.\ FWHM as measured with \sextractor for the optimal-resolution mosaic (top) and the deepest-lower-resolution mosaic (bottom). We excluded detections with magnitude errors $>$ 0.3 mag, \sextractor flags $>$ 3, and FWHM $\leq$ $0\farcs69$ for the optimal-resolution and FWHM $\leq$ $0\farcs99$ for the optimal-depth mosaic. The solid lines are the FWHM-limit used in each image, and the dotted-dash line is the surface-brightness limit. Red points indicate stars selected based on their magnitude and FWHM.  \label{figure:magrad}}
\end{figure}

\begin{figure}
\centerline{
\includegraphics[width=0.48\txw]{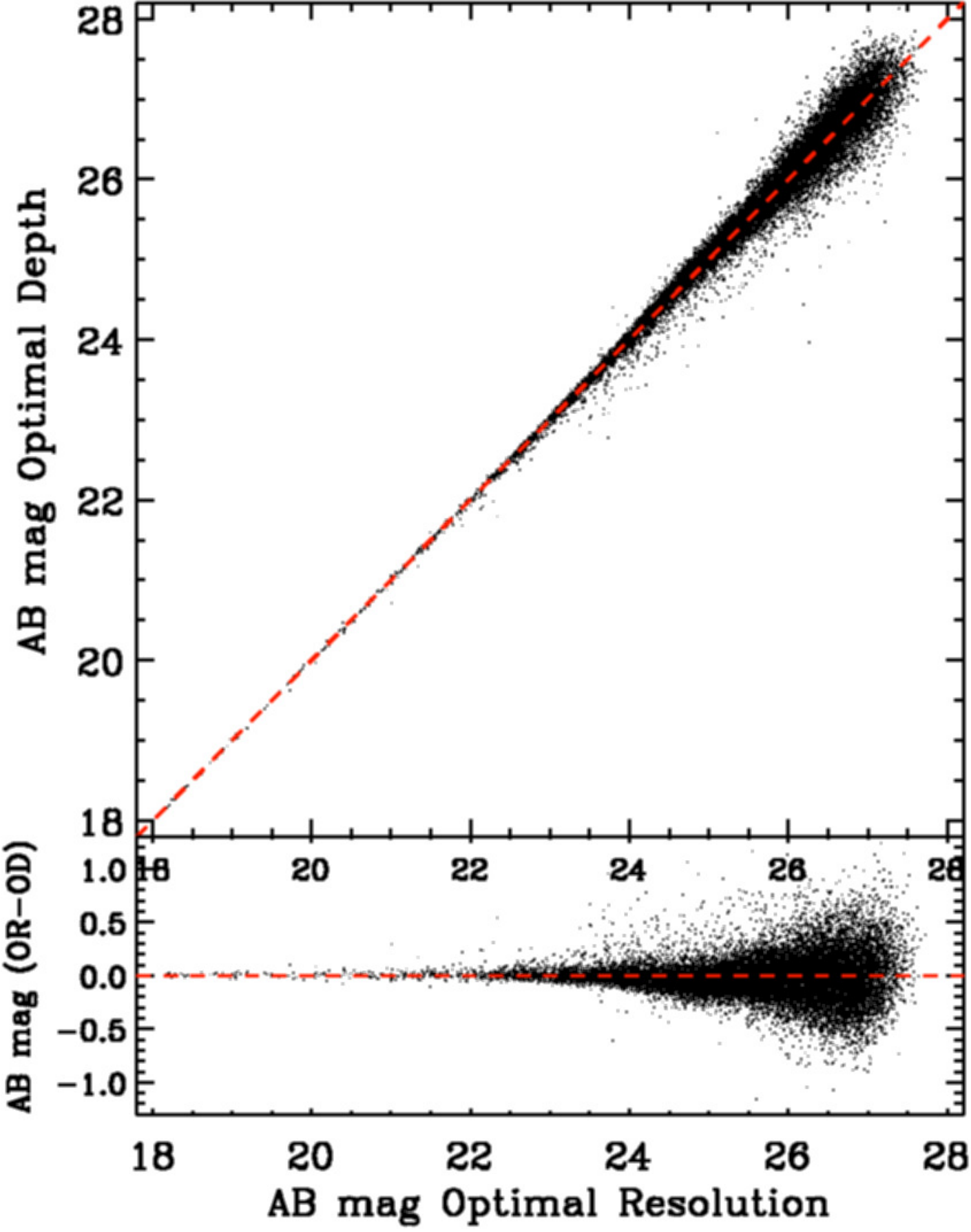}
}
\caption{ \noindent\small
Comparison of total \Ub magnitudes measured by \sextractor in the optimal-resolution (OR) and the optimal-depth (OD) mosaics. In the bottom panel, we subtracted the \Ub total magnitude of the optimal-depth mosaic from that measured in the optimal-resolution mosaic.  \label{figure:mvsm}}
\end{figure}

\begin{figure}
\centerline{
\includegraphics[width=0.48\txw]{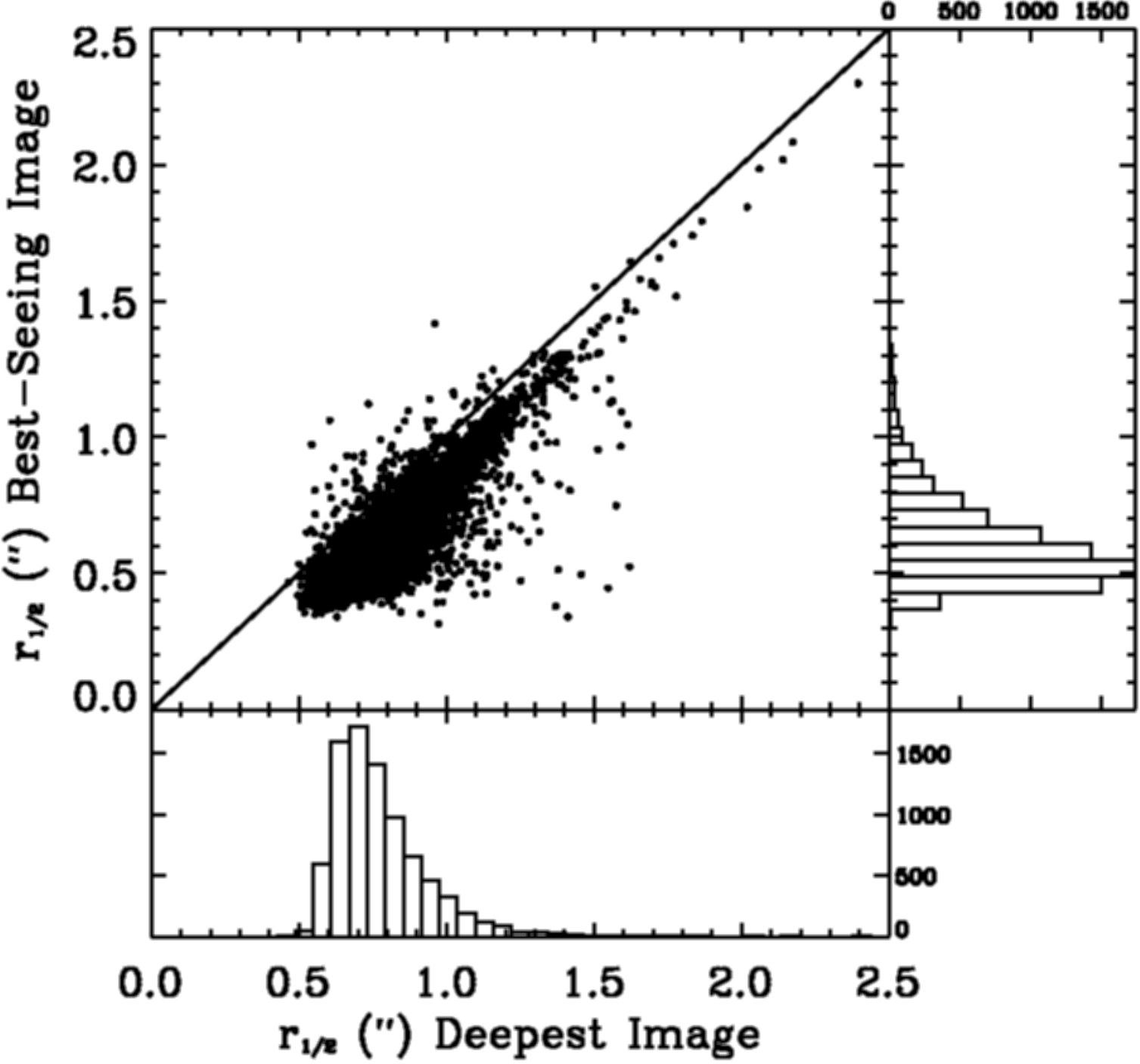}
}
\caption{ \noindent\small
Half-light radius for galaxies with \Uab $\lesssim$26 mag were measured using SExtractor. The optimal-resolution image indeed yields consistently smaller half-light radii compared to the optimal-depth image, although the best fit regression still has a slope close to unity: r$_{\textrm{optimal-resolution}}$ $\simeq$ 1.00 r$_{\textrm{optimal-depth}}-0\farcs15$. \label{figure:rvsr}}
\end{figure}

\subsection{Image Depths and Completeness of \Ub Mosaics}

Fig. \ref{figure:magrad} compares the object magnitude versus the half-light radius measured by \sextractor for the optimal-resolution image (top) to the lower-resolution image (bottom). The dot-dashed line represents the surface brightness limit for each of the mosaics.

We randomly inserted $10^{3}$ artificial point sources into each mosaic to characterize the actual point source detection limits. The resulting 5$\sigma$ limit for each mosaic are presented in Table \ref{table:stack}. The lower-resolution deepest image is 90$\%$ and 50$\%$ complete at \Uab$\lesssim 27.0$ mag and \Uab$\lesssim 28.0$ mag respectively, while the shallower optimal-resolution stack is  90$\%$ and 10$\%$ complete to the same magnitude limits. All image stacks, including the deepest image, begin to deviate from 100$\%$ completeness at fluxes fainter than \Uab $\gtrsim  25.5$ mag. This drop-off in completeness is more gradual for the deepest-lower-resolution mosaics, but is more dramatic for the highest resolution mosaics.

\subsection{Optimal-Resolution vs.\ Optimal-Depth LBT \Ub Mosaics}
The optimal-resolution and optimal-depth catalogs were matched, and in Fig.\,\ref{figure:mvsm}, the total magnitude measured by \sextractor for each object in both mosaics are compared. Fig.\,\ref{figure:mvsm}b shows agreement in total magnitude within 0.5 mag for the majority of objects out to \Uab$\sim$27 mag.  The half-light radius measured by \sextractor for both the optimal-resolution and optimal-depth images for galaxies brighter than \Uab$\lesssim$ 26 mag is shown in Fig.\,\ref{figure:rvsr}. The optimal-resolution image consistently measures smaller half-light radii compared to the optimal-depth image with the half-light radii histogram peak being $\sim0\farcs5$ and $\sim0\farcs7$, for the optimal-resolution image and the optimal-depth image, respectively (Fig.\,\ref{figure:rvsr}).

There are clear advantages and disadvantages in excluding the data acquired during poorer seeing conditions from our final mosaics of the GOODS-N field. In the optimal-resolution stack, sub-structures (e.g., knots) within the brightest and largest galaxies are more pronounced and discernible (Fig.\,\ref{figure:brightgals336} -- Fig.\,\ref{figure:brightgals3}), making them better suited for studies of their morphology.

Besides the drop-off in the resulting galaxy counts fainter than \Uab$\simeq$\hspace{0.1cm}26 mag, another disadvantage of the optimal-resolution mosaic is the loss of low-surface brightness emission in the outer parts of faint galaxies. In general, this does not seem to be a significant effect (Fig.\,\ref{figure:brightgals336} -- Fig.\,\ref{figure:brightgals3}), and should not prevent one from using the optimal-resolution stacks for galaxies as faint as \Uab$\simeq$ 25.5 mag in total flux, and to surface brightness levels of \sbab = 32 mag arcsec$^{-2}$ in our GOODS-N LBT \Ub images, as discussed in \S$\,$\ref{sec:sbprofiles}.

To detect the faintest distant galaxies (\Uab = 28.1 mag), the deepest images possible are required that include  almost all usable \Ub data. The middle two images of Fig.\,\ref{figure:faintgals} highlight the additional fainter galaxies detected by \sextractor in the optimal-depth mosaic.  Comparing the optimal-depth LBT \Ub image (Fig. \ref{figure:faintgals} middle-right) to the HST \emph{B}--band (F435W; Giavalisco et al. 2004) and \emph{U}--band (F336W; \textit{HST} Program: 13872 PI: Oesch) images (Fig. \ref{figure:faintgals} far-left and far-right, respectively) confirms that the faintest detected galaxies in the \Ub image are in fact real.

To create the redshift distributions in Fig.\,\ref{figure:histz}, redshifts for GOODS-N were taken from the 3D HST catalog (Skelton et al.\,2014) and include photometric and spectroscopic redshifts. Photometric redshifts were determined with the EAZY code by fitting the spectral energy distribution (SED) composed of photometric data covering the $0.3$\textendash$8 \mu$m wavelength range (Skelton et al. 2014). When available, spectroscopic redshifts were used from the literature as summarized by Skelton et al.\,(2014). Using our LBT \Ub optimal-resolution and optimal-depth catalogs, histograms of object redshifts (Fig. \ref{figure:histz}) show that most objects have redshifts $0\lesssim z\lesssim 3$, and that more objects are detected in the deepest LBT mosaic compared to the shallowest but optimal-resolution LBT mosaic. The ratio of detected galaxies between the optimal-resolution and optimal-depth catalogs is consistent for most redshifts, and only slightly increasing with increasing redshift. The highest redshift galaxies detected in the \Ub in this field (2.5 $\lesssim \, z \, \lesssim \,$3) are also the smallest and faintest galaxies detected. As the redshift increases, the average size of the galaxies sampled generally decreases (e.g., Ferguson et al. 2004; Windhorst et al. 2008). Despite the increase in depth in the lower-resolution mosaic (of the image including nearly all \Ub exposures), the smaller galaxy sizes and the PSF-FWHM $\geq 1\farcs0$ inhibits our ability to detect a larger fraction of all galaxies at the very faintest flux levels (\Uab $>$ 27) and at highest redshifts. Moreover, the rapid decline at $z\gtrsim$ 2.5 further reflects that at these redshifts, the \Ub filter begins to sample well below Ly$\alpha$ and the 912$\angstrom$ Lyman limit, where very few objects emit any significant light (e.g. Mostardi et al. 2013, Smith et al. 2016; Grazian et al. 2017).

\subsection{\Ub Weighted Image Stacks}

In \S$\,$\ref{sec:mosaics},, we described how the individual images were re-sampled and co-added with \textsc{swarp}. These images were median-combined, and their FWHM was only used for selection purposes.  To try to minimize the impact of these larger FWHM images while preserving image depth, we next combined the images while taking into account a weight-factor based on the FWHM of \textit{each individual} image. The weight-map for each image was multiplied by the inverse square of the FWHM in arcseconds (i.e., $ W*$FWHM$^{-2} $). An image with a FWHM of $1\farcs0$ would have a weight of 1. Using these modified weight maps, we re-ran \swarp the same way, except for the parameter ''combine\_type", but changed from ``median" to ``weighted". The formula for weighted co-addition of the images is: \begin{equation} F =  \frac{\sum w_if_i}{\sum w_i}, \; \; \; \; \textrm{where} \; \;  w_i  = \frac{W_i}{(\textrm{FWHM})_i^2} \end{equation}

\begin{figure}[!]
\centerline{
\includegraphics[width=0.51\txw]{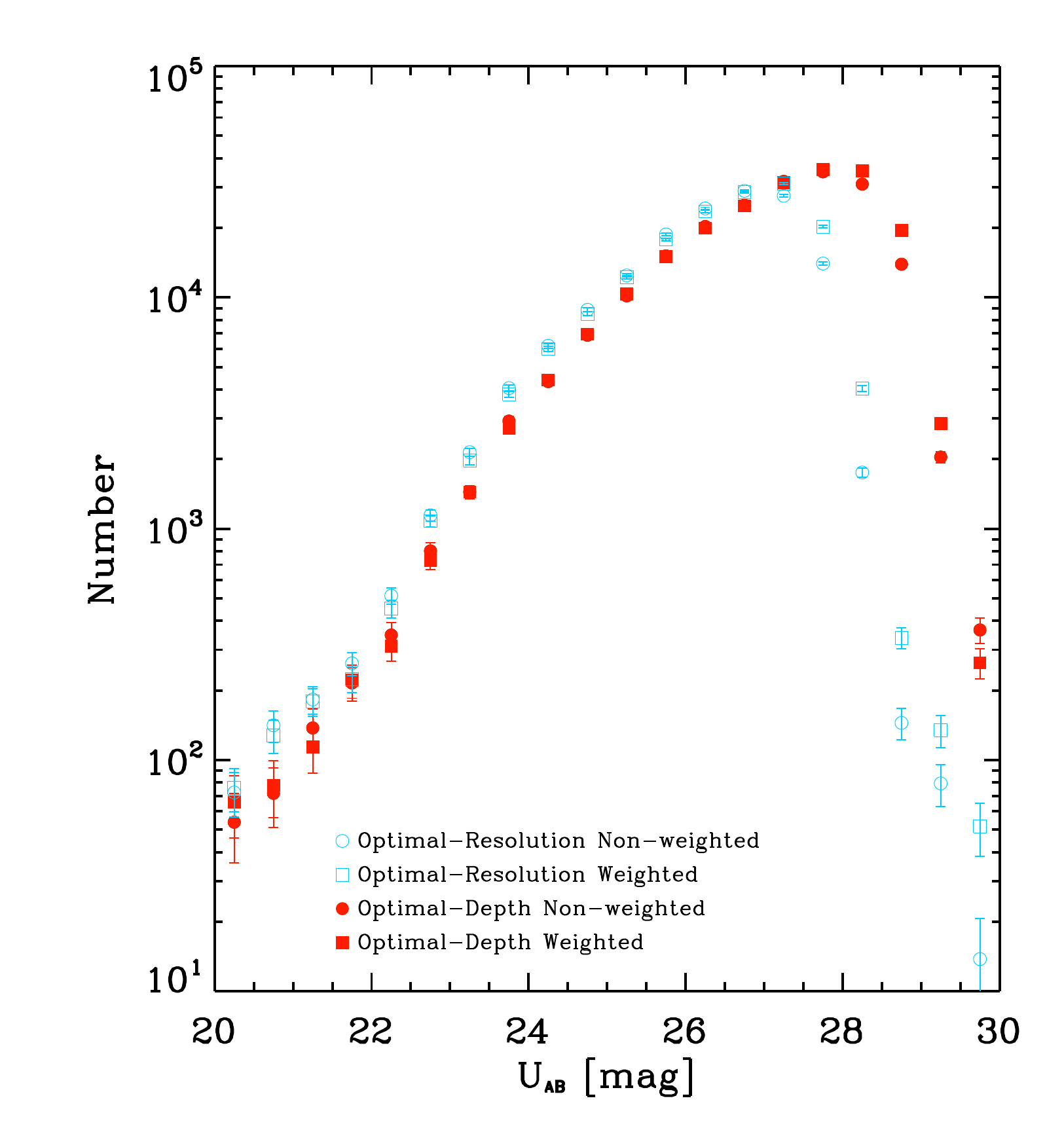}
}
\caption{\noindent\small
Number counts for both the median-combined stack (circles) and weighted-stacks (squares) for the optimal-resolution (blue) including all images with FWHM $\lesssim0\farcs8$ and the deepest mosaic (red) including all images with FWHM $\lesssim1\farcs8$. For the optimal-resolution and optimal-depth mosaics, the weighted and median stacks show the same trend until the incompleteness decline sets in around \Uab$\gtrsim27$ mag and \Uab$\gtrsim28$ mag, respectively. At the faintest magnitudes, the weighted image detects more objects.   \label{figure:numcntswht}}
\end{figure}

Compared to the unweighted image stacks, the weighted \Ub stacks detect a slightly larger number of faint objects in both the optimal-depth and the optimal-resolution mosaics (Fig. \ref{figure:numcntswht}) compared to the unweighted mosaics. For the shallower optimal-resolution image, the FWHM of the stars does not show a noticeable change, while for the deeper-image the average FWHM for the stars decreased from $1\farcs1$ to $1\farcs0$. This shows that when combining data of widely varying seeing FWHM-values --- as is usually the case in ground-based image stacking --- proper weighting needs to be applied to each image seeing.

\subsection{\Ub Surface Brightness Profiles of Well-Resolved Galaxies within GOODS-N}
\label{sec:sbprofiles}
\begin{figure*}[t!]
\centerline{
  \includegraphics[width=1.05\txw]{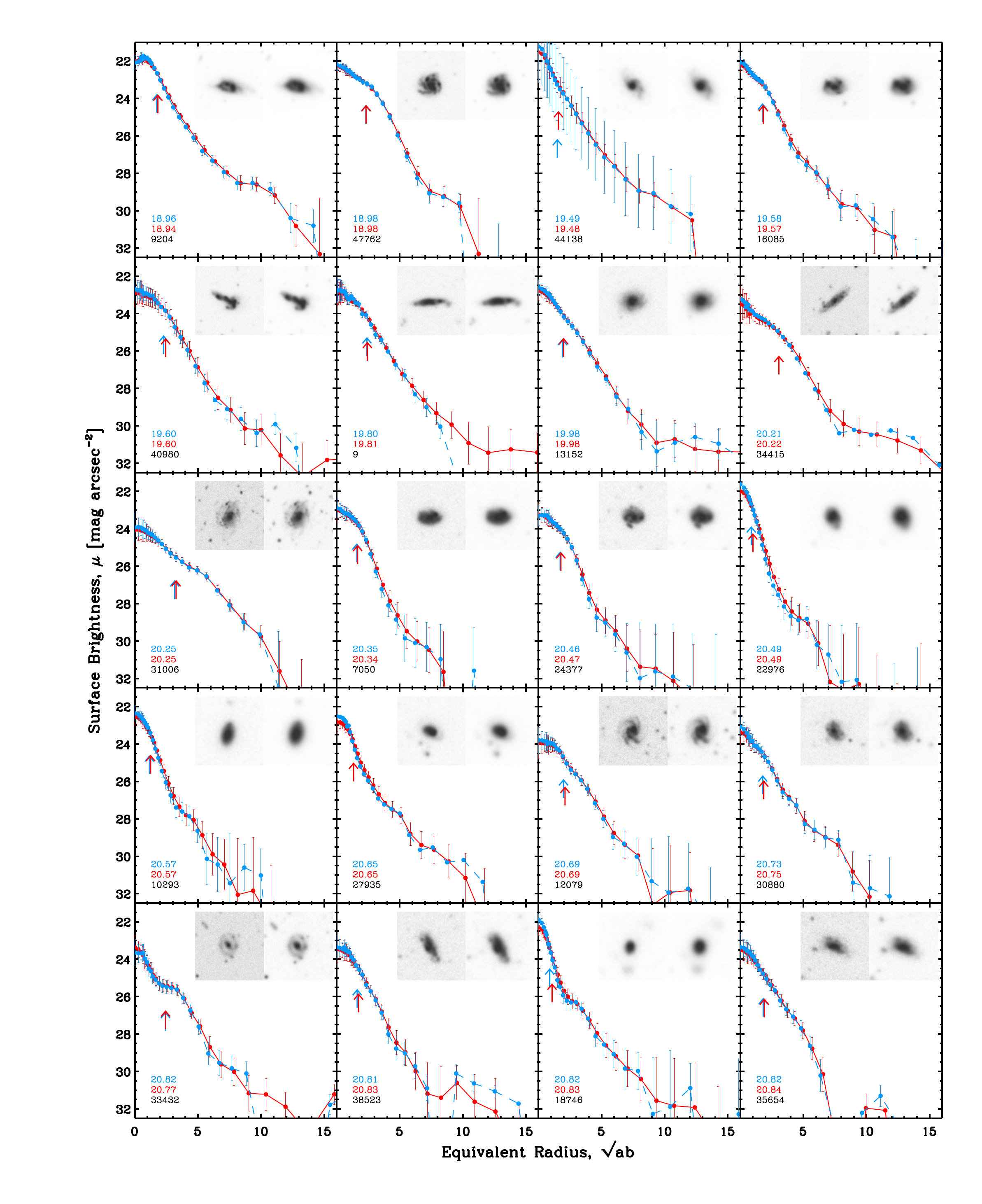}
}
\caption{\noindent\small
Surface brightness profiles for the 220 brightest objects with \Uab$\lesssim 23$ mag. Blue data are the optimal-resolution stacks, and red data are the optimal-depth stacks. The blue dashed line is the light-profile for optimal-resolution image and the red solid line is for the deepest-lower-resolution mosaic. Total \sextractor \Uab mags are also shown in each bottom left corner in blue/red, and the black number is the \sextractor catalog number. The blue/red arrows represent the half-light radius as measured by \sextractor. The left insert image is the optimal-resolution and the right image includes the optimal-depth image.    \label{figure:sbp}}
\end{figure*}

We selected the 220 brightest (AB\,$\lesssim$ 23 mag) and most extended galaxies, and measured their azimuthally averaged radial surface brightness (SB) profiles for both the optimal-resolution and the lower-resolution stacks. The majority of the galaxy sample are face-on and edge-on spirals, and the remainder are mostly early-type galaxies. We did so using a custom \idl procedure \texttt{galprof}\footnote{http://www.public.asu.edu/~rjansen/idl/galprof1.0/galprof.pro} written by one of us (RAJ), which performs surface photometry within a set of growing elliptical annuli. \sextractor segmentation maps were used as input to \texttt{galprof} to separate galaxy and background pixels.  Fig. \ref{figure:sbp} includes the sample of all 220 galaxies with \Uab $\lesssim$ 23 mag in order of decreasing flux. For each galaxy, the figure includes the SB-profile and corresponding grey-scale images from both the optimal-resolution and the optimal-depth mosaics. For the majority of the galaxies, the SB-profiles are similar with only very subtle differences, to within the $1\sigma$ SB-profile errors. The optimal-resolution SB-profile generally starts off slightly brighter in the center with the SB-profile dropping off slightly faster than the deepest image SB-profile. There are a few exceptions to this, which are also shown in Fig \ref{figure:sbp}. 

\noindent
\vspace*{-\baselineskip}

To compare the results that \texttt{galprof} outputs, we plot the total \Ub magnitudes measured by \texttt{galprof} in the optimal-resolution and the optimal-depth mosaics for the 220 galaxies in Fig.\,\ref{figure:sbmodels} (left panels). The bottom panel of Fig. \ref{figure:sbmodels} shows the \Ub total magnitude of the optimal-depth mosaic subtracted from that measured in the optimal-resolution mosaic versus the magnitude of optimal-resolution. For galaxies fainter than \Uab$\simeq$ 21 mag, there is an offset in total magnitude with the optimal-resolution image having a brighter total flux. This offset is only 0.05 mag and may be due to over-subtraction of the background in the optimal-depth mosaic. To test our results from \texttt{galprof}, we made model galaxies with sersic profiles and varied input parameters to match the variety of galaxies in our 220 sample and measured them using \texttt{galprof}. The range of input properties are: 1) AB magnitude from 19-23 mag, 2) half-light radius (r$_{e}$) from 0.9"-2.48", and 3) axial ratio from 0.1-1.0. Each model galaxy was convolved with the PSF corresponding to the optimal-resolution or optimal-depth image. The PSFs were produced by averaging 25 stars which were not saturated and at least 800 pixels away from the border. In Fig. \ref{figure:sbmodels} (right) the total \Ub magnitudes measured by \texttt{galprof} are shown for the optimal-resolution and the optimal-depth mosaics for the model galaxies. Our model galaxies are overall consistent with the real galaxies, but the slight bias toward the optimal-resolution mosaic remains visible for the rare large galaxies at the fainter magnitudes.

\begin{figure*}
\plottwo{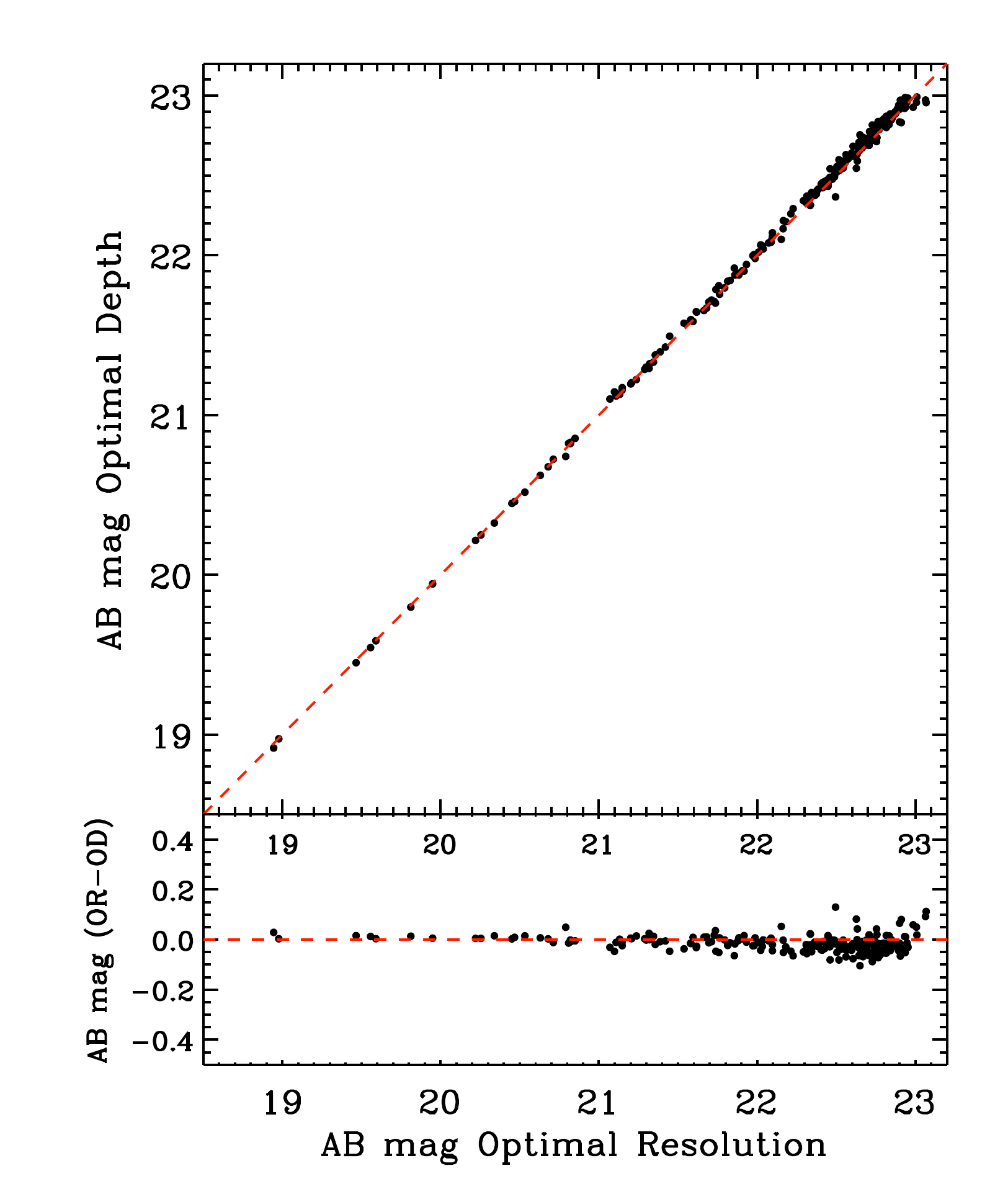}{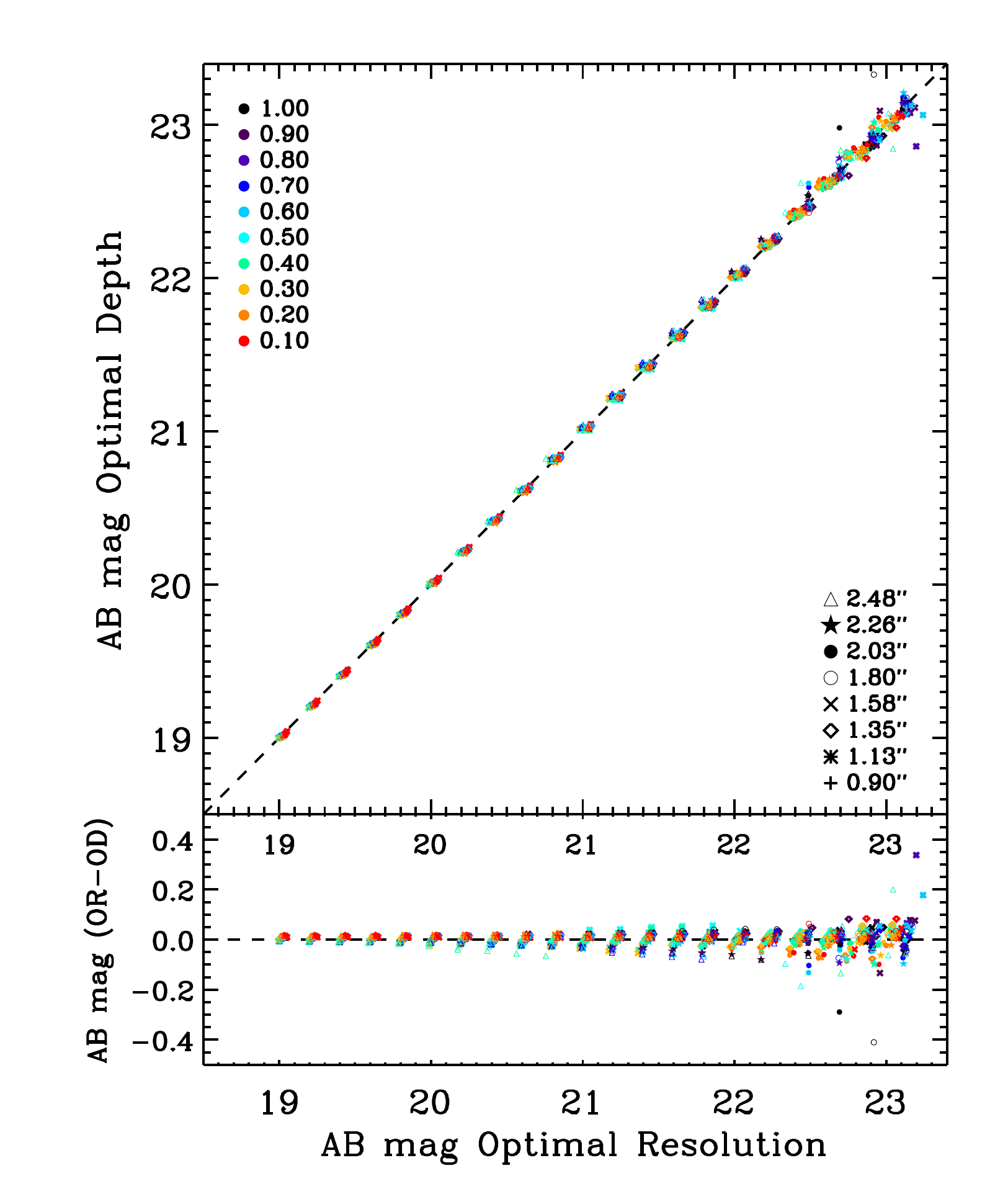}
\caption{ \noindent\small
(Left) Comparison of total \Ub magnitudes measured by \texttt{galprof} in the optimal-resolution and the optimal-depth mosaics for the 220 brightest galaxies in the field. (Right) Comparison of total \Ub magnitudes measured by \texttt{galprof} in the optimal-resolution and the optimal-depth mosaics for model galaxies artificially inserted into the image. The different colors represent the different b/a axis ratios used and the different symbols represent the different half-light radius. (Both) In the bottom panel, we subtracted the \Ub total magnitude of the optimal-depth mosaic from that measured in the optimal-resolution mosaic.  \label{figure:sbmodels}}
\end{figure*}

\subsection{Implications for the Extragalactic Background Light}
This results is important in the context of potentially large amounts of missing light in the outskirts of galaxies that have been mentioned as a possible explanation of the high values of the \emph{direct} extragalactic Background Light (EBL) values in the literature (e.g. Bernstein, Freedman, \& Madore 2002). Recent results by Driver et al. (2016) used very deep panchromatic galaxy number-count data to estimate the integrated extra-galactic background light (iEBL) in 20 different filters from the far-UV to the far-IR. The counts in all 20 filters were deep enough that they converged with well determined faint-end slopes, so that the sky-integral of the integrated EBL could be determined to within acceptable errors (10--20\%, see Fig. 3 of Driver et al. 2016), which were modeled with Monte Carlo simulations. Driver et al. (2016) found significantly smaller iEBL values, by factors 3--8, in the optical-blue to the near-IR compared to the \emph{direct} EBL measurements from various sources (for a review, see Dwek \& Krennrich 2013). They argue that this discrepancy could be due to foreground light sources (Zodiacal light and the Milky Way galaxy) possibly not having been fully subtracted from the direct EBL measurements, so that the iEBL method that uses the galaxy number-counts may be seeing most of the real EBL. Our uniquely deep LBT \Ub imaging allows us to determine if the Driver et al. (2016) results could still be underestimating the true EBL from the integrated galaxy counts, due to significant missing light hiding in the low-surface brightness outskirts of galaxies (Dwek \& Krennrich 2013 and references therein). 

In this exercise, we only look at the brightest galaxies, because they dominate the EBL total energy budget in the universe at \mab$\simeq$20--23 mag, simply because in the optical the largest change in count-slope occurs in this flux range, from non-converging at brighter magnitudes to well converging at much fainter magnitudes (Windhorst et al. 2011, Driver et al. 2016). As a consequence, about half of the \Ub EBL power comes from the flux range \mab$\simeq$20--23 mag, which is precisely the range where our LBT light-profiles in Fig. \ref{figure:sbp} do not show a large amount of missing flux in the galaxy outskirts when comparing our more sensitive lower-resolution images to the less sensitive highest-resolution images. Examining the SB-profiles of our 220 brightest galaxies, with \Uab$\lesssim 23$ mag, we found that fewer than 20 galaxies (or $\lesssim$10$\%$) show more than a $\gtrsim 0.5$ mag difference in the SB-profile outskirts to \Uab$\lesssim 32$ mag arcsec$^{-2}$. This is also shown in Fig. \ref{figure:mvsm}b, which does not show a large systematic flux difference between the optimal resolution and optimal depth images to \Uab$\lesssim$ 23 mag, which corresponds to no more than 0.1 mag for the entire population. 
There appears to be not enough low-surface brightness emission in the outskirts of the brighter galaxies to explain the large (factor 3--5) difference between the two methods of computing the EBL. Hence, it is unlikely that the blue iEBL derived from the integrated counts is missing a large amount of low-SB emission in galaxy out-skirts. Fig. \ref{figure:sbp} simply shows that an insufficient amount of light is hiding (in the \Ub filter) in the outskirt of galaxies to explain the significant discrepancy between the direct blue EBL measurements and the integrated EBL values of Driver et al. (2016). 

One caveat is that we can only do this currently in the U-band, because this is the LBT filter for which we have a largest number of exposures available that cover a wide range in seeing. Redder filters would be more sensitive to any missing galaxy bulge or halo-light. Another caveat is that our study cannot constrain or rule out truly diffuse sources of EBL as a possible cause of the above discrepancy. Such sources are, e.g., inter-group or inter-cluster light, or truly unresolved intergalactic populations, which possibilities are discussed in Driver et al. (2016). In conclusion, bright galaxies (\Uab $\lesssim$ 23 mag) that are known to produce most of the EBL, do \emph{not} seem to be missing more than $\sim$ 0.05 -- 0.10 mag of their total light in galaxy outskirts on $\gtrsim$ $1\farcs0$ scales to \Uab $\lesssim$ 32 mag arcsec$^{-2}$.

\section{DISCUSSION AND SUMMARY}
\label{sec:summary}
Typical \Ub seeing at the LBT as measured from stars in LBC images is $\sim$ $1\farcs0-1\farcs1$ FWHM, and usually worse for the \Ub (Taylor et al. 2004). The current study combines exposures taken on many different nights with varying atmospheric seeing conditions with the telescope observing the \textit{same} part of the sky. While HST needs 15 separate pointings to cover the GOODS-N field, the large FOV of the LBC encompasses it in just one. We used 315 separate \Ub exposures of the GOODS-N field to explore and compare mosaicing the best-seeing subset of images to mosaicing all usable images. At \Uab $\gtrsim 26$ mag, our optimal-resolution image no longer detects the same number of galaxies as the deepest, lower-resolution image. The drop-off in the number counts is more dramatic for the shallower optimal-resolution image, and more gradual for the full and deeper stack of all usable images.

We conclude that for studies of brighter galaxies and features visible within them, the optimal-resolution image should be utilized. However, to fully explore and understand the faintest objects the deepest imaging with the lower-resolution is required, as it gives better sensitivity to lower-surface brightness objects.

By weighting the images based on the quality of the FWHM when stacking, we are able to improve the FWHM of the final \Ub  mosaic, \emph{and} detect a larger number of fainter objects. We recommend such weighting when co-adding LBC and other ground-based images, especially when the images have a wide range in FWHM. This method does not add significant amount of processing time to the stacking procedure, and is already a feature in the \swarp program.

From the ground in the \U-band, we are able to reach $\sim0\farcs8$ resolution FWHM and detect isolated objects to $\sim28$ mag. These ground-based images will never be able to compete with HST for resolution ($0\farcs07$-$0\farcs09$ in F336W, see e.g. Windhorst et al. 2011), which is needed to do pixel to pixel analysis. For photometry measurements the main challenge is overcoming the confusion limit for separating objects which occurs once objects are closer than about $\sim1\farcs0$. The advantage of well observed fields like GOODS-N is the availability of the HST \B-band. With the addition of HST \B-band, packages like ConvPhot (De Santis et al. 2007) or T-fit (Laidler et al. 2007) can separate objects within the LBT confusion limit to measure the flux associated with individual objects as determined by HST resolution.

For the 220 brightest galaxies with \Uab$\lesssim$ 23 mag, we measured the surface brightness profiles in both the optimal-resolution and optimal-depth mosaics. By comparison there is only marginal differences between the light-profiles to \sbab$\lesssim$ 32 mag arcsec$^{-2}$. In only 10\% of the cases are the total-flux differences larger than 0.5 mag. This helps constrain how much flux can be missed in galaxy outskirts, which is important for studies of the Extragalactic Background Light. Trujillo \& Fliri (2016) imaged a nearby galaxy in $r$-band (UGC 00180) and obtained radial surface brightness limit $\sim$ 33 mag arcsec$^{-2}$ with the 10.4 m Gran Telescopio de Canarias telescope. They found only $\sim$ 3\% of total light to be in the stellar halo which agrees with theoretical predictions.

Our sub-stacking method can easily be implemented on imaging with the LBT and other telescopes. Even when the LBT transitions to ``Queue-observing" for all large programs, getting sub-arcsecond seeing for the entire program will be nearly impossible, particularly at the shortest wavelengths (\U--band). Making mosaics while only stacking the best-seeing subset of images is therefore one way to fully utilize the potential of these unique data sets.

For future surveys with limited observing time, in the age of queue observing, a requirement of sub-arcsecond seeing is the only way to fully take advantage of these large telescopes and optimize the science. Such data will however be very hard to obtain in the \U--band and will require many nights of observing.

\acknowledgments

Acknowledgements. The LBT is an international collaboration among institutions in the United States, Italy, and Germany. LBT Corporation partners are The University of Arizona on behalf of the Arizona university system; Istituto Nazionale di Astrofisica, Italy; LBT Beteiligungsgesellschaft, Germany, representing the Max-Planck Society, the Astrophysical Institute Potsdam, and Heidelberg University; The Ohio State University; and The Research Corporation, on behalf of The University of Notre Dame, University of Minnesota, and University of Virginia. R. A. Windhorst acknowledges support from NASA JWST grants NAG-12460 and NNX14AN10G.

\facility{LBT(LBC)}.




\noindent\begin{figure}[h!]
\centerline{
  \includegraphics[width=0.96\txw]{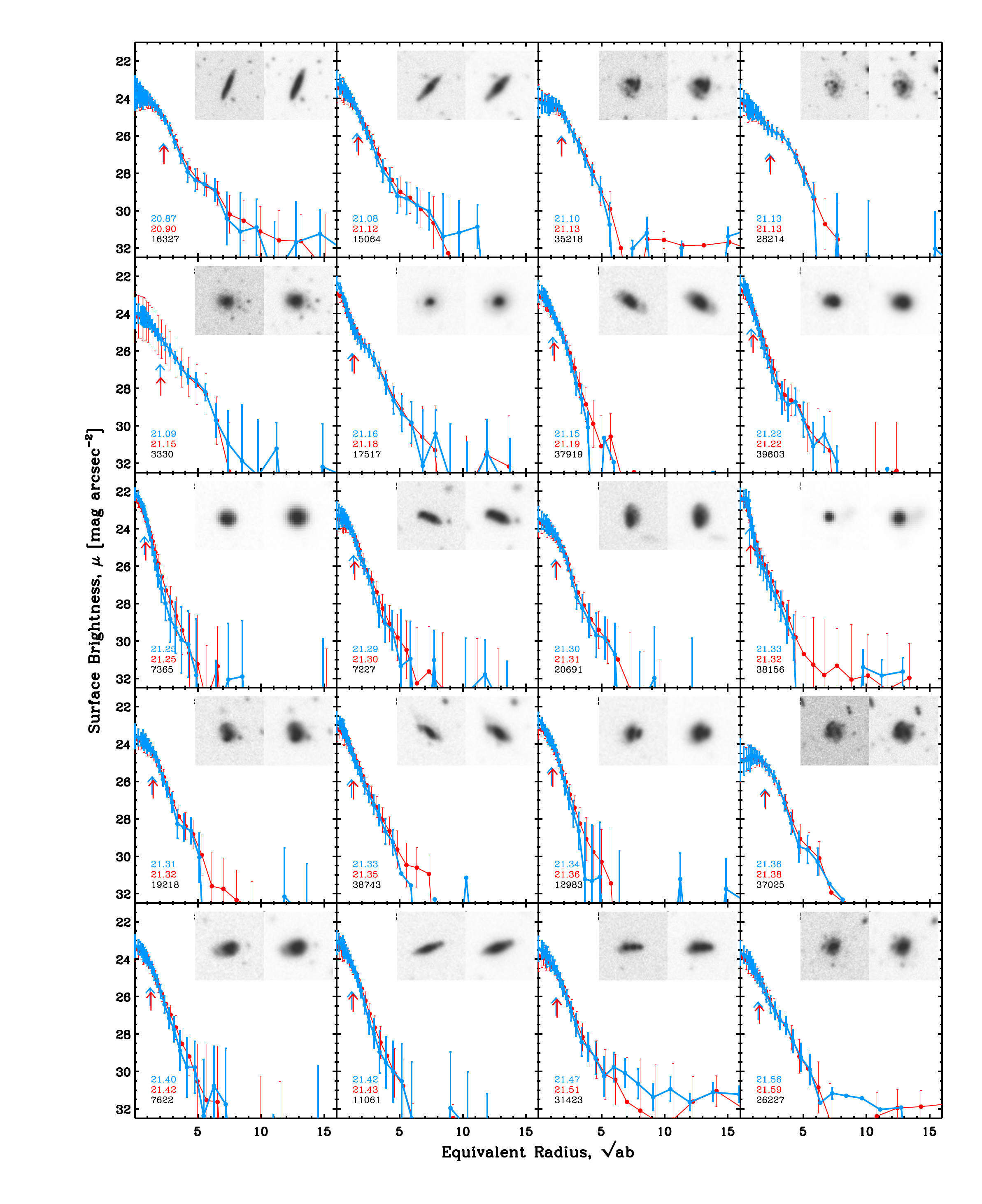}
}
\figurenum{13}
\caption{Cont. }
\end{figure}

\noindent\begin{figure*}[h]
\centerline{
  \includegraphics[width=0.96\txw]{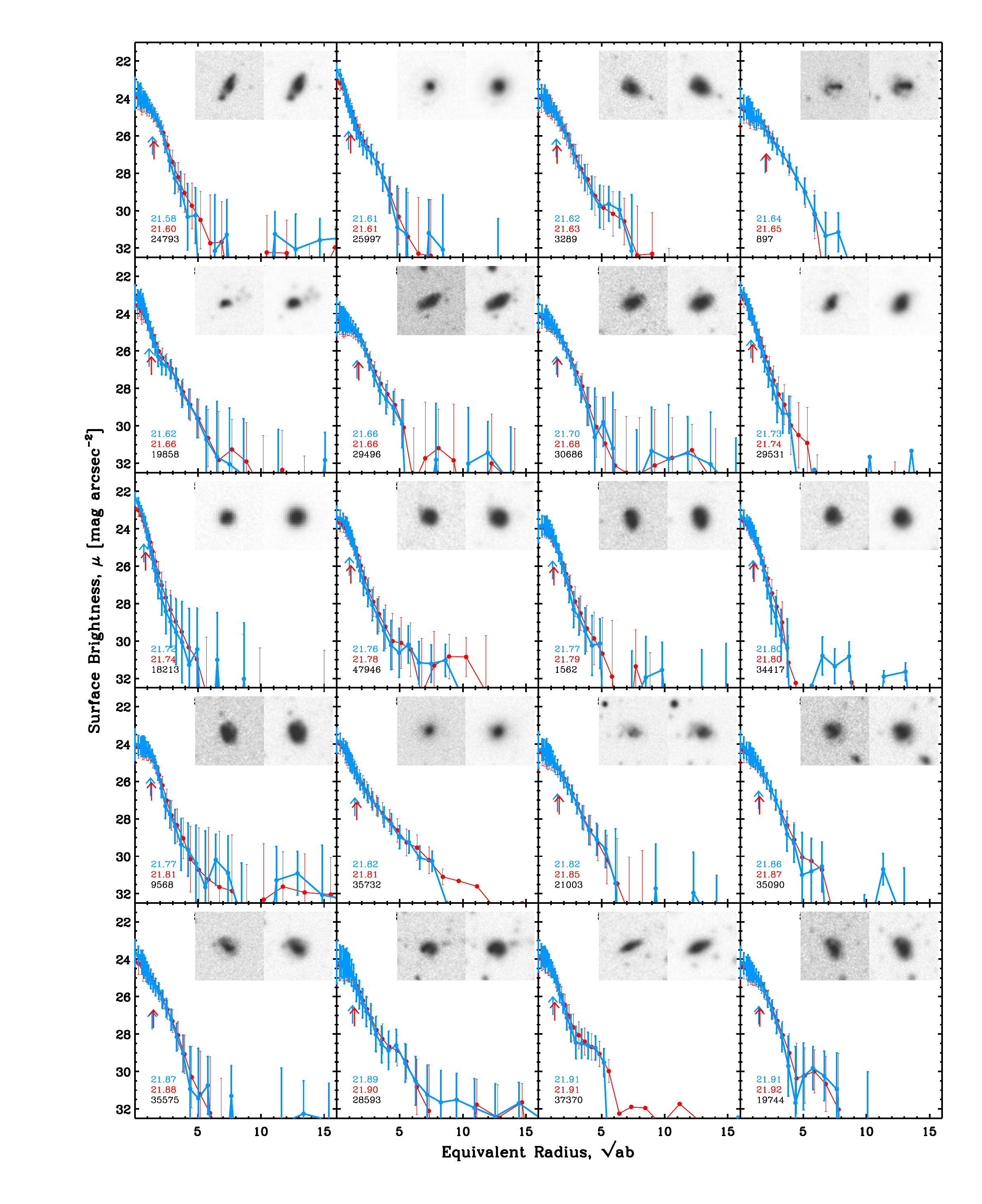}
}
\figurenum{13}
\caption{Cont. }
\end{figure*}

\noindent\begin{figure*}[h]
\epsscale{1.0}
\centerline{
  \includegraphics[width=0.96\txw]{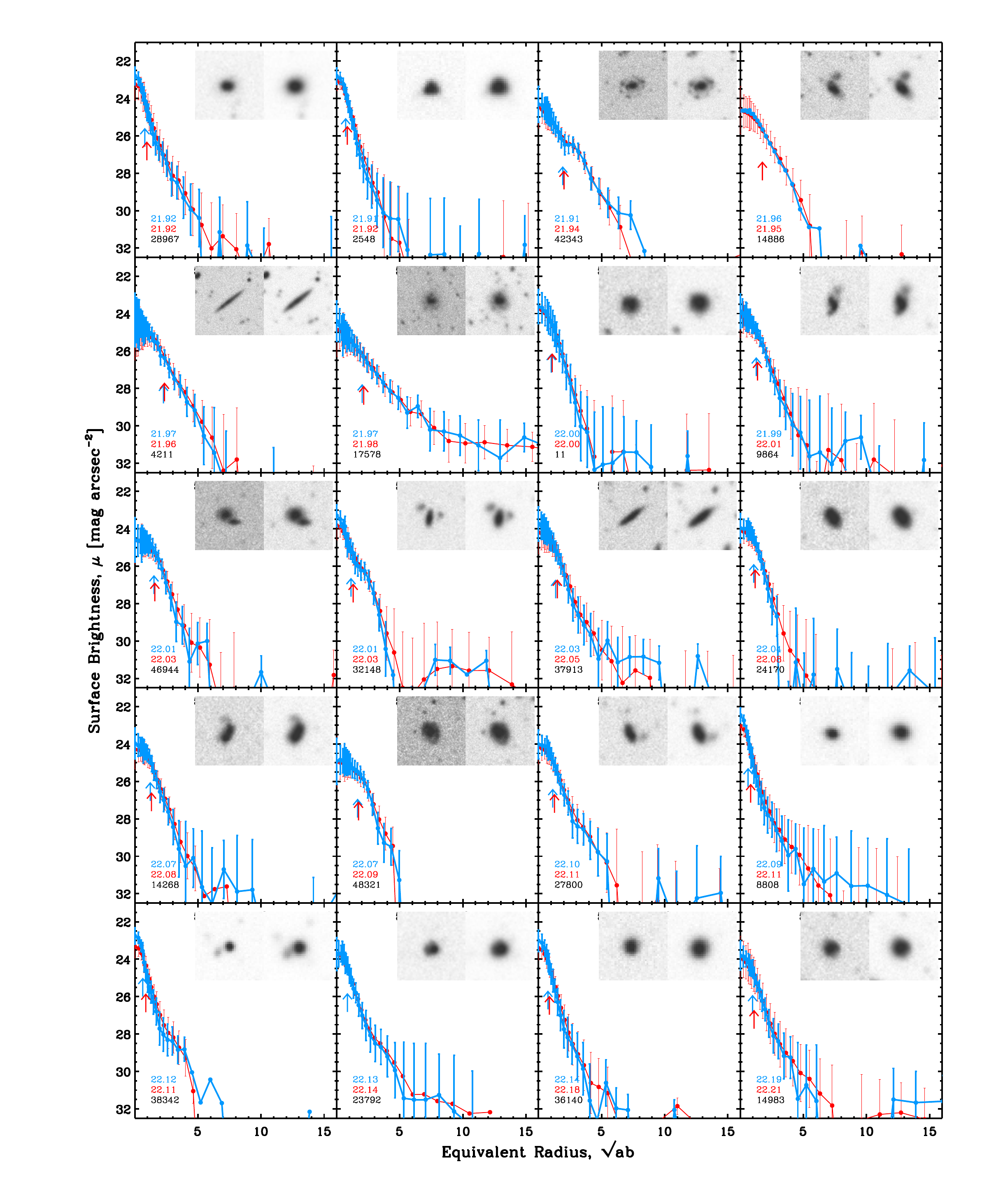}
}
\figurenum{13}
\caption{Cont. }
\end{figure*}

\noindent\begin{figure*}[h]
\epsscale{1.0}
\centerline{
  \includegraphics[width=0.96\txw]{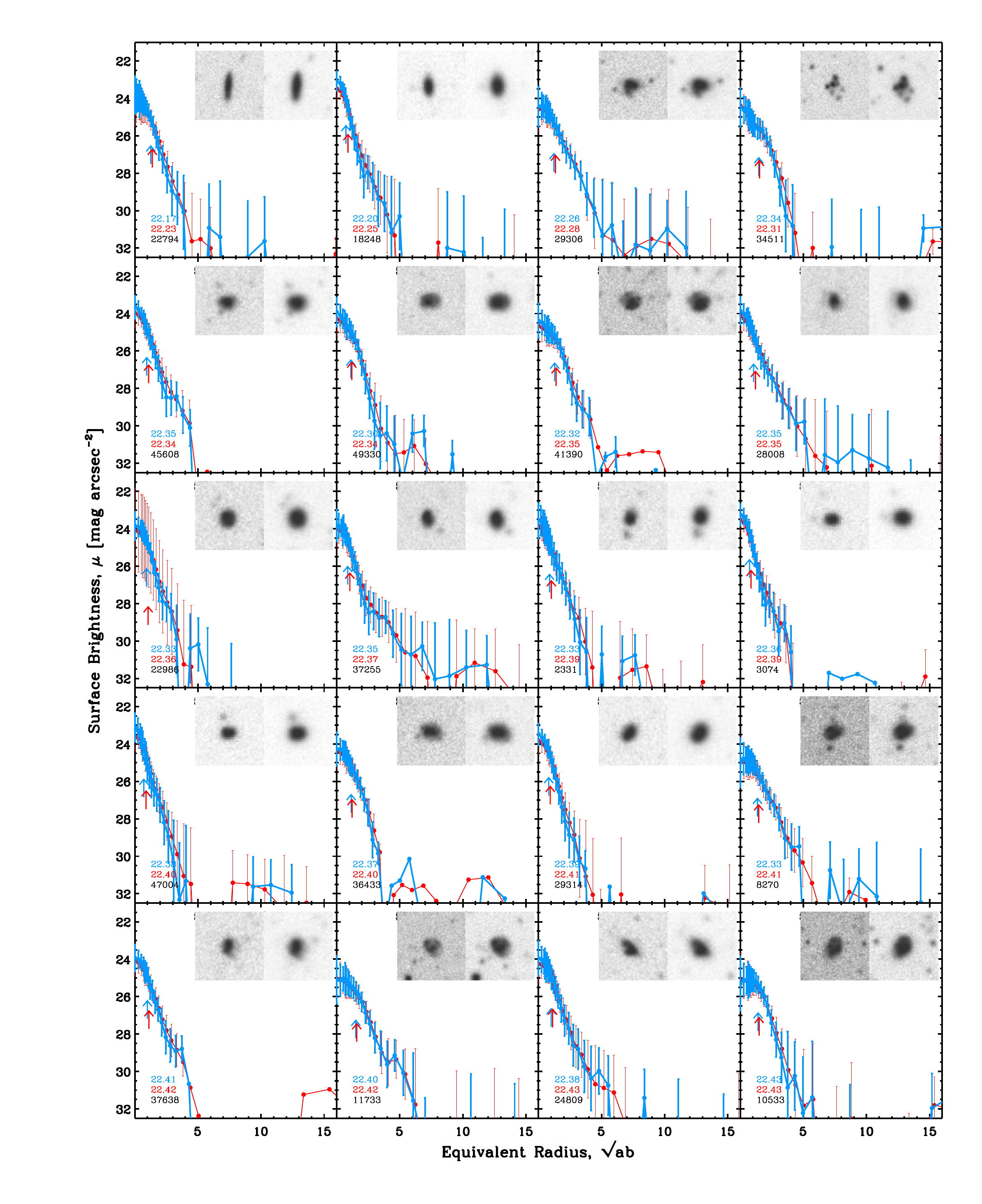}
}
\figurenum{13}
\caption{Cont. }
\end{figure*}

\noindent\begin{figure*}[h]
\epsscale{1.0}
\centerline{
  \includegraphics[width=0.96\txw]{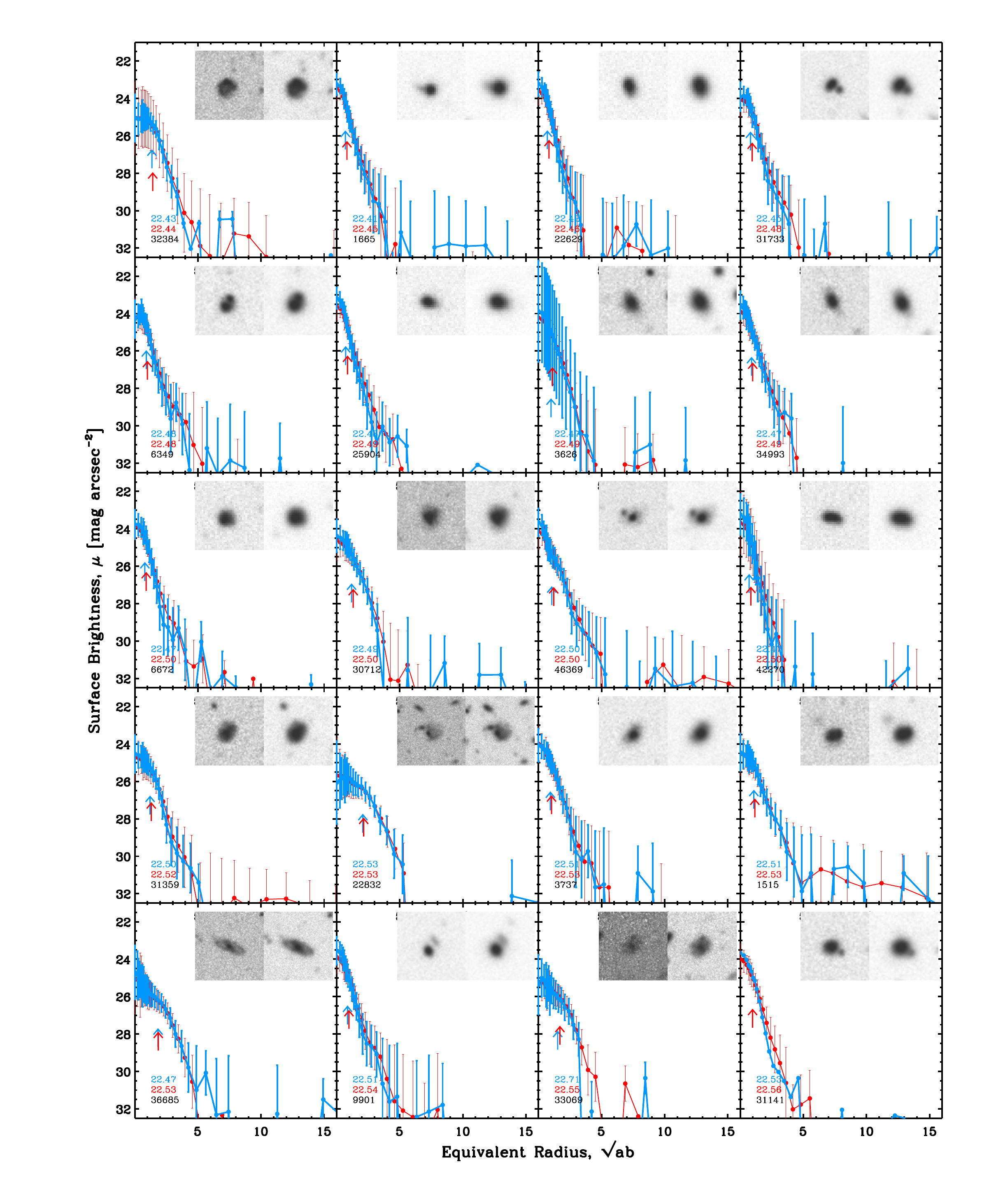}
}
\figurenum{13}
\caption{Cont. }
\end{figure*}

\noindent\begin{figure*}[h]
\epsscale{1.0}
\centerline{
  \includegraphics[width=0.96\txw]{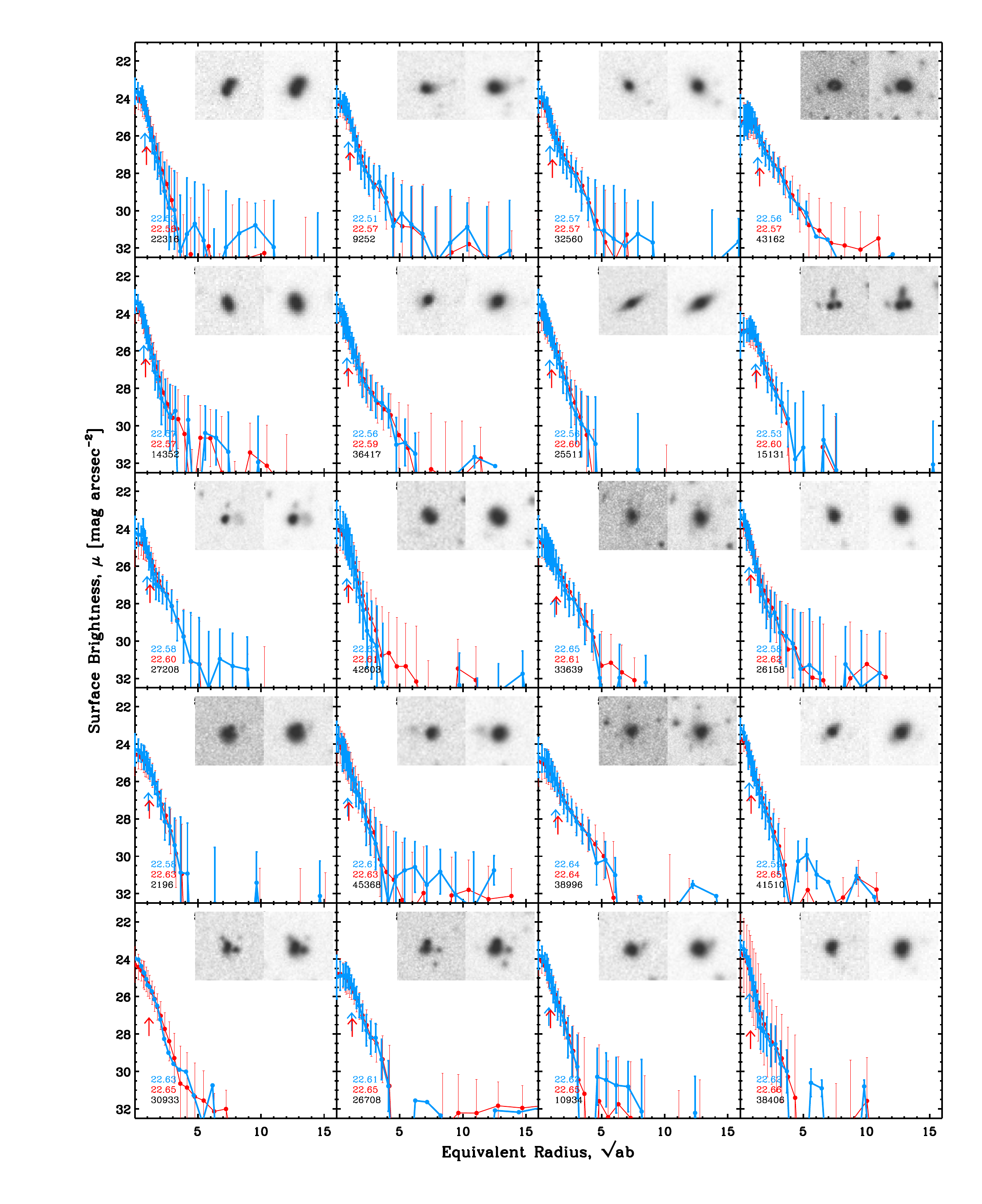}
}
\figurenum{13}
\caption{Cont. }
\end{figure*}

\noindent\begin{figure*}[h]
\centerline{
  \includegraphics[width=0.96\txw]{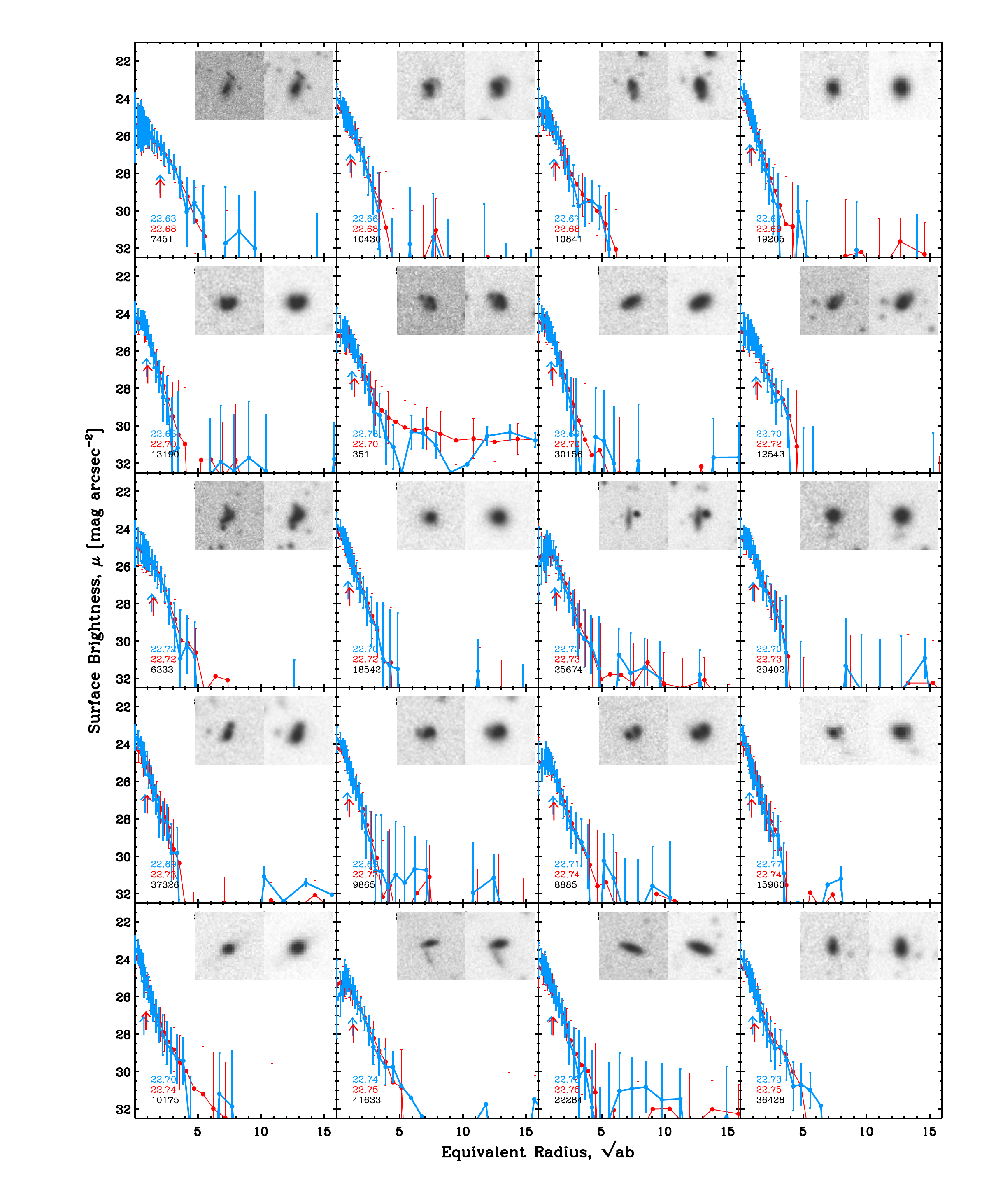}
}
\figurenum{13}
\caption{Cont. }
\end{figure*}

\begin{figure*}[h]
\centerline{
  \includegraphics[width=0.96\txw]{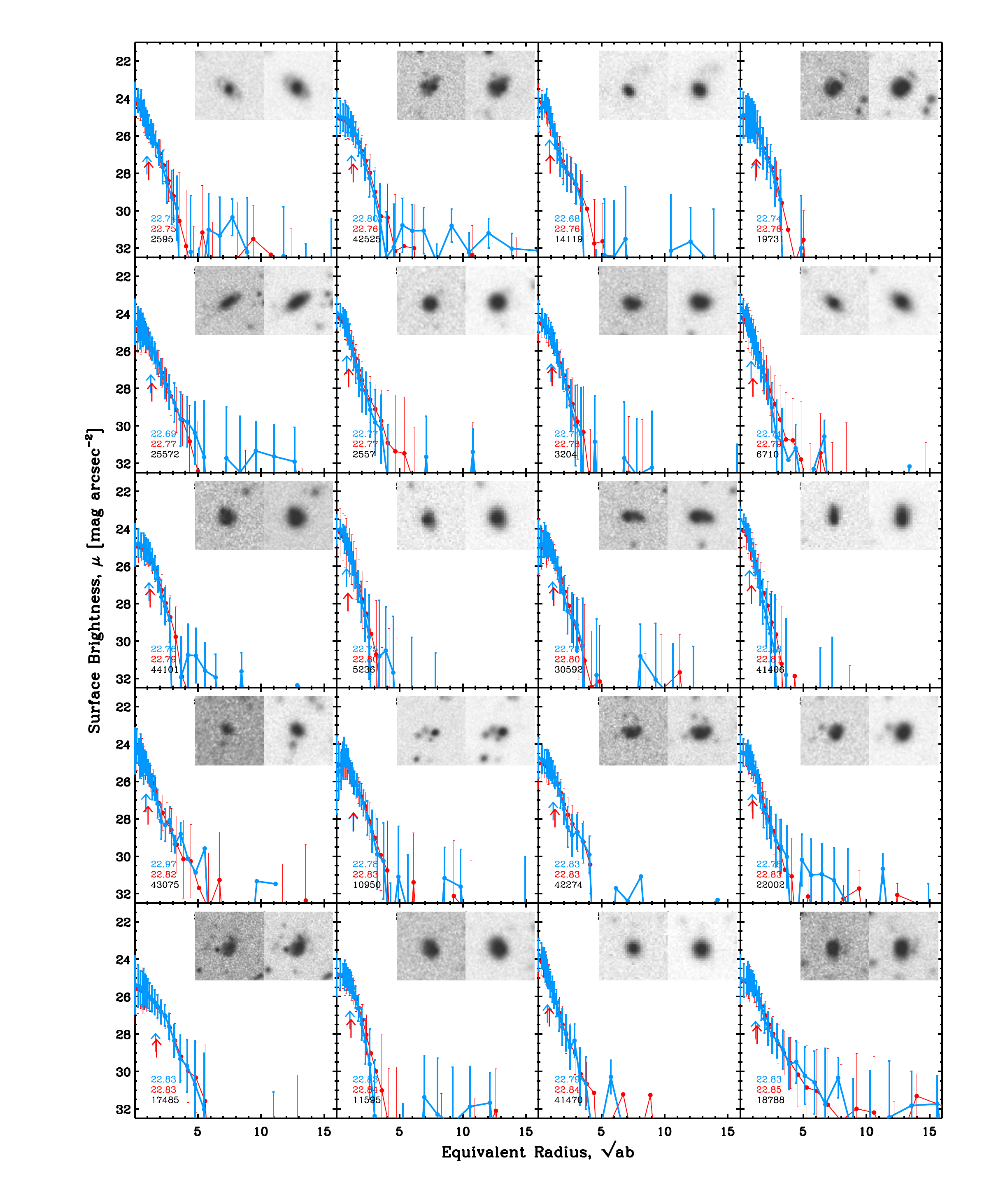}
}
\figurenum{13}
\caption{Cont. }
\end{figure*}

\noindent\begin{figure*}[h]
\epsscale{1.0}
\centerline{
  \includegraphics[width=0.96\txw]{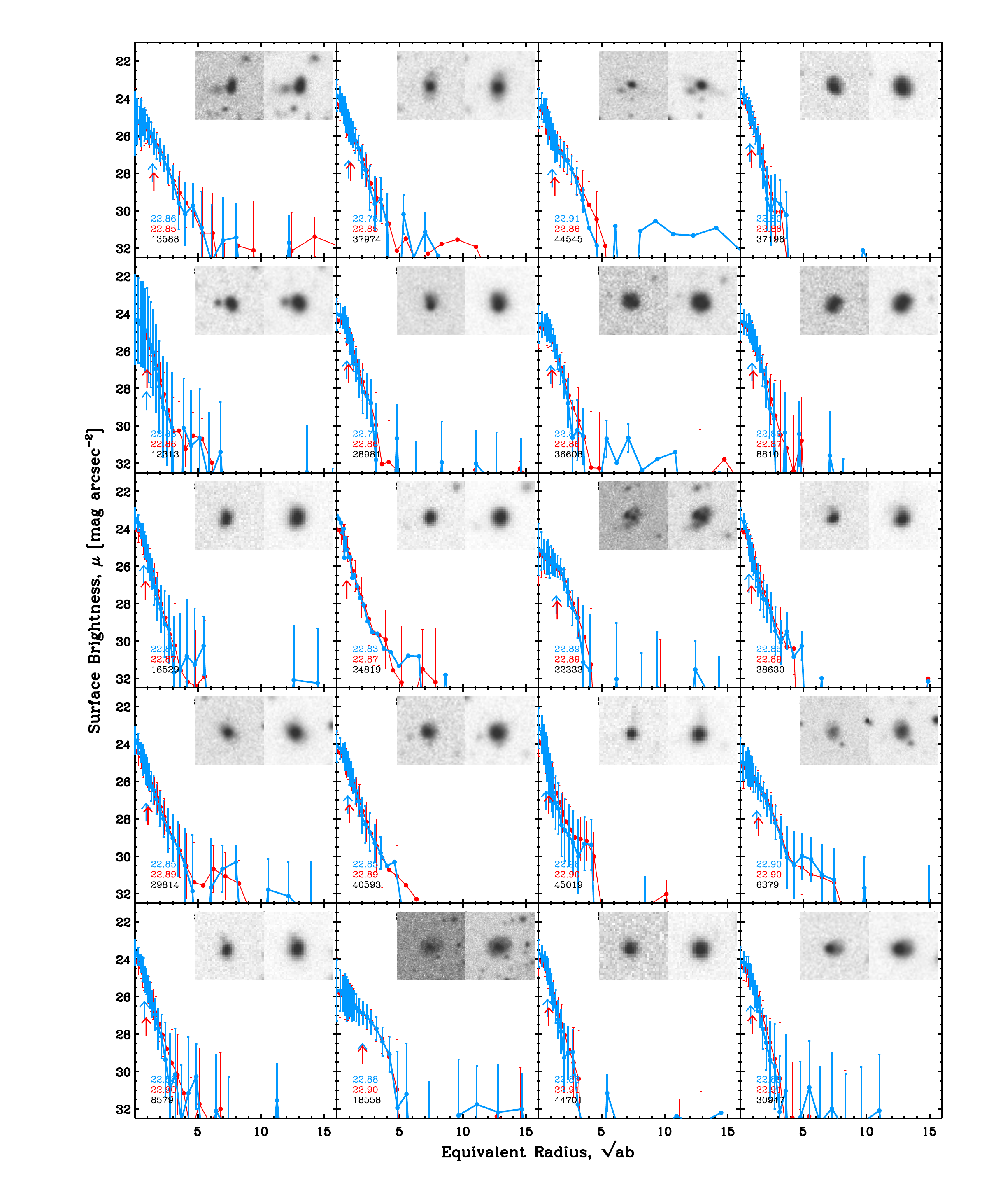}
}
\figurenum{13}
\caption{Cont. }
\end{figure*}

\noindent\begin{figure*}[h]
\epsscale{1.0}
\centerline{
  \includegraphics[width=0.96\txw]{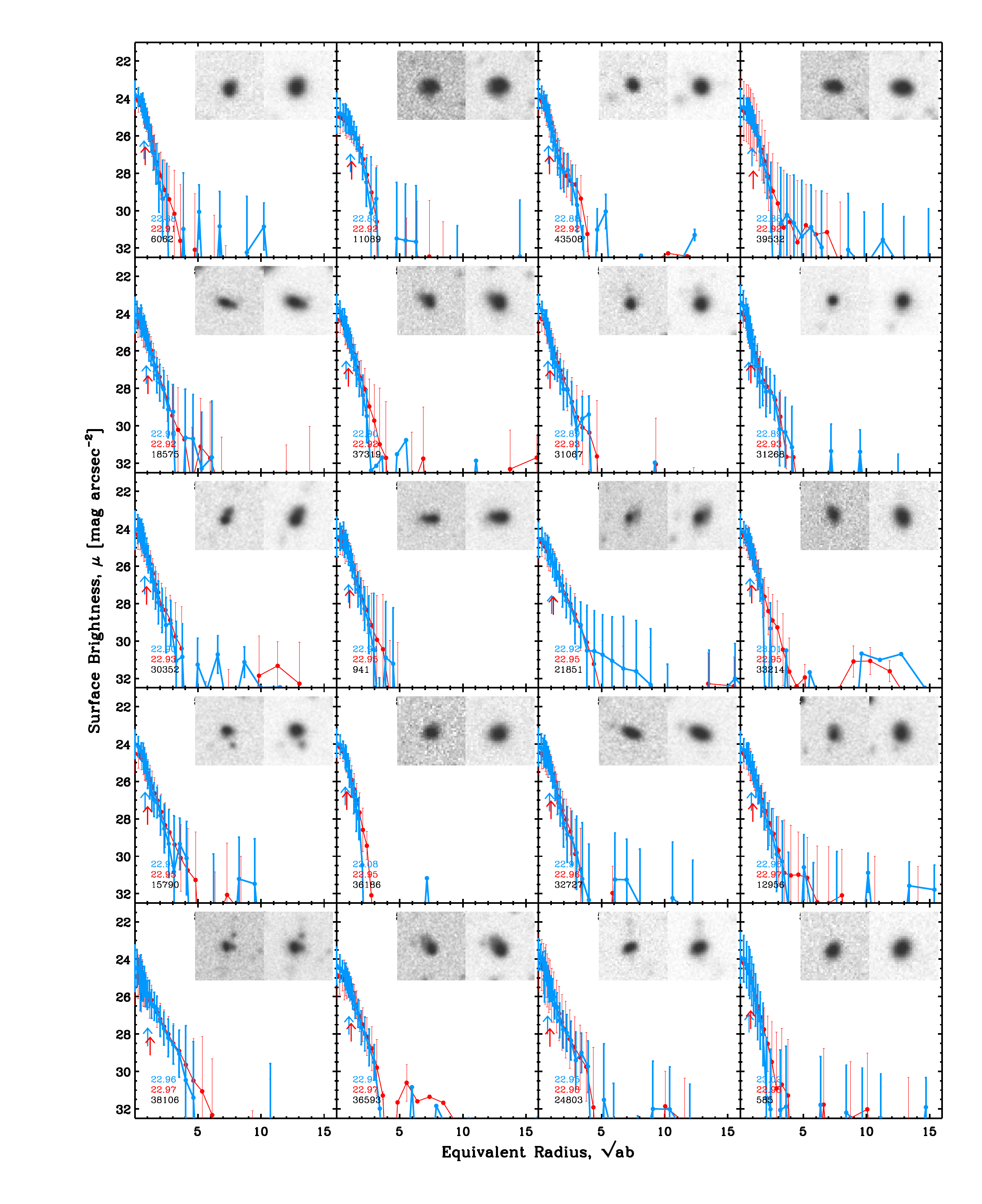}
}
\figurenum{13}
\caption{Cont. }
\end{figure*}

\end{document}